\documentclass{article}
\usepackage[utf8]{inputenc}
\usepackage{amsfonts,amsmath,amssymb,amsthm,graphicx,float,hyperref,color}
\usepackage{soul}

\definecolor{Blue}{rgb}{0,0,1}                  
\definecolor{Red}{rgb}{1,0,0} 
\definecolor{Green}{rgb}{0,1,0}    
\definecolor{Bronze}{rgb}{0.8,0.5,0.2}    
\definecolor{Violet}{rgb}{0.54,0.17,0.89}

\newcommand{\ab}{{\mathbf a}}
\newcommand{\bb}{{\mathbf b}}

\newcommand{\eb}{{\mathbf e}}

\newcommand{\ub}{{\mathbf u}}
\newcommand{\vb}{{\mathbf v}}

\newcommand{\p}{\partial}

\newcommand{\psib}{\boldsymbol{\psi}}
\newcommand{\nub}{\boldsymbol{\nu}}
\newcommand{\mub}{\boldsymbol{\mu}}
\newcommand{\phib}{\boldsymbol{\phi}}

\newcommand{\ba}{\begin{array}}
\newcommand{\ea}{\end{array}}
\newcommand{\be}{\begin{equation}}
\newcommand{\ee}{\end{equation}}
\newcommand{\bd}{\begin{displaymath}}
\newcommand{\ed}{\end{displaymath}}

\newcommand{\xb}{\mathbf{x}}

\newcommand{\C}{\mathbb{C}}
\newcommand{\R}{\mathbb{R}}

\newcommand{\YC}[1]{{#1 }}

\title{Quantum Physics without the Physics}
\author{N. Anders Petersson\thanks{Center for Applied Scientific Computing, Lawrence Livermore National Laboratory, PO Box 808, Livermore CA
94551. } \and Fortino Garcia\thanks{Department of Applied Mathematics, University of Colorado at Boulder, Boulder, CO 80309} \and Daniel E. A. Appel\"{o}\thanks{Department of Computational Mathematics, Science and Engineering and
Department of Mathematics, Michigan State University, East Lansing, MI 48824} \and Stefanie G\"unther$^*$ \and Youngsoo Choi$^*$ \and Ryan Vogt$^*$}

\date{December 1, 2020}

\begin{document}

\maketitle
\tableofcontents
\newpage
\section{Introduction}
This report explains the basic theory and common terminology of quantum physics without assuming any knowledge of physics. It was written by a group of applied mathematicians while they were reading up on the subject. The intended audience consists of applied mathematicians, computer scientists, or anyone else who wants to improve their understanding of quantum physics. We assume that the reader is familiar with fundamental concepts of linear algebra, differential equations, and to some extent the theory of Hilbert spaces. Most of the material can be found in the book by Nielsen and Chuang~\cite{Nielsen-Chuang} and in the lecture notes on open quantum systems by Lidar~\cite{lidar2020lecture}. Another excellent online source of information is Wikipedia, even though most of its articles on quantum physics assume a solid understanding of physics.

The fundamental postulates and theory of quantum physics are introduced in Sections~\ref{sec_notation}-\ref{sec_dens-post}. The dictionary part of the report is contained in Sections~\ref{sec_closed} and ~\ref{sec_open}. Here, short definitions of commonly used terminology are presented, both in terms of state vectors (Section~\ref{sec_closed}) and in terms of density matrices (Section~\ref{sec_open}).

\section{Notation}\label{sec_notation}
This document uses a mix of matrix-vector and Dirac bra-ket notation. In matrix-vector notation, column vectors are set in boldface lower case symbols, e.g. $\psib$ or $\vb$. Upper case letters denote matrices (operators),
but we follow the convention in quantum physics and denote the density matrix by $\rho$. The
Hermitian conjugate (conjugate transpose) of a matrix $A$ is denoted by $A^\dagger$ and the
Hermitian conjugate of a column vector $\vb$ is the row vector $\vb^\dagger$.

Dirac bra-ket notation~\cite{Nielsen-Chuang} is often used in the quantum physics literature. It is related to
matrix-vector notation through
\[
| v\rangle \equiv \vb,\quad \langle v | \equiv \vb^\dagger,\quad 
\langle u | A | v \rangle \equiv
\langle \ub, A \vb\rangle = \ub^\dagger A \vb,
\quad \langle u | v \rangle \equiv
\langle \ub,\vb\rangle = \ub^\dagger \vb.
\]
Here the standard $\ell_2$ scalar product and norm for vectors $\ub$ and $\vb$ in $\mathbb{C}^d$ are are defined by
\begin{equation}\label{eq_sp+norm}
  \langle \ub, \vb \rangle = \sum_{j=0}^{d-1} \bar{u}_j v_j,\quad \| v\| = \sqrt{\langle \vb, \vb \rangle}.
\end{equation}

The terms (linear) operator and matrix are used interchangeably, referring to a
linear mapping (morphism) between two vector spaces. We remark that there is no ambiguity in this notation because we only consider finite dimensional vector spaces. In that case, all linear operators can be represented by a matrix once a basis has been selected.

\section{Schr\"odinger's equation as a partial differential equation}
In its most general form, the Schr\"odinger equation governing the evolution of a quantum state $\Psi$ (also called a wave function) is
\begin{equation} \label{eq_hamiltonian-abstract}
  i \hbar \frac{\p}{\p t} \Psi = \hat{H} \Psi.
\end{equation}
Here, $t$ is time, $\hbar=h/2\pi$ is the reduced Plank constant and $\hat{H}$ is a Hamiltonian
operator. Depending on the application, the Hamiltonian operator can take many different forms. For example, a single particle with mass $m$ in one space dimension can be modeled by
\[
\hat{H} = \frac{\hat{P}^2}{2 m} + V(x,t),\quad \hat{P} = -i\hbar \frac{\p}{\p x},
\]
where $x$ is the Cartesian coordinate, $\hat{P}$ is the momentum operator and $V(x,t)$ is a scalar real-valued function describing the potential energy.
This Hamiltonian leads to the 1-D Schr\"odinger equation,
\begin{equation}
  i \hbar \frac{\p}{\p t} \Psi = -\frac{\hbar^2}{2m} \frac{\p^2}{\p x^2}\Psi + V(x,t)\Psi.
\end{equation}
This equation can be generalized to model $N$ particles in 3-D, leading to a time-dependent partial differential equation (PDE) in $3N$ spatial dimensions. Due to the high dimensionality, the PDE formulation of such problems would be extremely challenging to solve numerically.

\subsection{Quantum harmonic oscillator}
Many concepts of quantum mechanics can be illustrated by the modeling of a quantum harmonic oscillator. In this case, the potential function is of the form,
\[
V(x) = \frac{1}{2}m\omega^2 x^2,
\]
where $\omega$ is the angular frequency of the oscillator,
leading to Schr\"odinger's equation with the Hamiltonian operator
\[
\hat{H}\Psi = -\frac{\hbar^2}{2 m}\frac{\p^2}{\p^2 x}\Psi + \frac{m\omega^2}{2} x^2\Psi,
\]
We define the scaled momentum and position operators by
\[
\hat{p}\Psi = -i x_0 \frac{\p}{\p x}\Psi,\quad \hat{q}\Psi = \frac{x}{x_0} \Psi,\quad x_0 = \sqrt{\frac{\hbar}{m\omega}},
\]
which allows the Hamiltonian operator to be expressed as
\[
\hat{H}\Psi = \frac{\hbar\omega}{2}\left(\hat{p}^2 + \hat{q}^2\right){\Psi}.
\]
Because
\[
\hat{p}\hat{q}\Psi = -i x_0 \frac{\p}{\p x}\left(\frac{x}{x_0} \Psi\right) = -i\Psi + \hat{q}\hat{p}\Psi,
\]
we have
\[
i\left( \hat{q}\hat{p} - \hat{p}\hat{q}\right) \Psi = - \Psi.
\]
Therefore,
\[
\left( \hat{q} - i\hat{p}\right)
\left( \hat{q} + i\hat{p}\right)\Psi = 
\left(
\hat{q}^2 + \hat{p}^2 - \hat{I}
\right)\Psi
\]
where $\hat{I}$ is the identity operator. This leads to the factored form of the Hamiltonain,
\begin{equation}\label{eq_ladder}
\hat{H}\Psi = \hbar\omega\left( \hat{a}^\dag \hat{a} + \frac{1}{2}\right)\Psi,\quad
\hat{a}^\dagger = \frac{1}{\sqrt{2}}\left( \hat{q} - i \hat{p}\right),\quad
\hat{a} = \frac{1}{\sqrt{2}}\left( \hat{q} + i \hat{p}\right).
\end{equation}
Here, $\hat{a}$ is called the lowering (annihilation) operator. The adjoint operator, $\hat{a}^\dagger$, is called the raising (creation) operator. 

The factored form of the Hamiltonian operator allows its eigenfunctions and eigenvalues to be calculated in an elegant manner. First note that the function
\begin{equation}\label{eq_ground-state}
 \phi_0(x) = \alpha e^{-x^2/2 x_0^2},\quad \alpha = \frac{1}{\pi^{1/4} x_0^{1/2}},
 \quad 
 \langle\phi_0, \phi_0 \rangle := \int_{-\infty}^\infty |\phi_0(x)|^2\, dx = 1,
\end{equation}
satisfies $\hat{a}\phi_0 = 0$. Thus, $\phi_0$ is a normalized eigenfunction of $\hat{H}$,
\[
\hat{H} \phi_0 = \lambda_0 \phi_0,\quad \lambda_0 = \frac{\hbar \omega}{2}.
\]
The eigenvalue $\lambda_0$ represents the energy level of the eigenfunction $\phi_0(x)$. In the quantum physics literature $\phi_0(x)$ is called the ground state and is denoted by $|0\rangle$.

It is straightforward to show the commutation relation between the lowering and raising operators,
\begin{equation}\label{eq_commute}
[\hat{a}, \hat{a}^\dagger] := \hat{a}\hat{a}^\dagger - \hat{a}^\dagger \hat{a} = \hat{I},\quad\Leftrightarrow\quad
\hat{a}\hat{a}^\dagger = \hat{a}^\dagger \hat{a} + \hat{I}.
\end{equation}
Because of the factored form of the Hamiltonian in \eqref{eq_ladder}, this leads to
\[
\hat{H}\hat{a}^\dagger = \hat{a}^\dagger \hat{H} + \hbar\omega \hat{a}^\dagger,\quad \Leftrightarrow \quad
[ \hat{H}, \hat{a}^\dagger] = \hbar\omega\hat{a}^\dagger.
\]
The function $\hat{a}^\dagger \phi_0$ thus satisfies 
\[
\hat{H} \hat{a}^\dagger \phi_0 = \hat{a}^\dagger \hat{H}\phi_0 + \hbar\omega \hat{a}^\dagger\phi_0 = \hbar\omega\left(1 + \frac{1}{2} \right) \hat{a}^\dagger \phi_0,
\]
and is an non-normalized eigenfunction of $\hat{H}$ with eigenvalue $\lambda_1 = \hbar\omega(1 + 0.5)$.
In general, the (non-normalized) eigenfunctions of $\hat{H}$ are of the form $\left(\hat{a}^\dagger\right)^n\phi_0$ because repeated use of \eqref{eq_commute} gives $\hat{a}^\dagger \hat{a}\left(\hat{a}^\dagger\right)^n\phi_0 = n \left(\hat{a}^\dagger\right)^n\phi_0$. Based on this property, we define the number operator by 
\begin{equation}\label{eq_number-op}
  \hat{n} = \hat{a}^\dagger \hat{a},\quad \hat{n} \left(\hat{a}^\dagger\right)^k\phi_0 = k \left(\hat{a}^\dagger\right)^k\phi_0,\quad k=0,1,2,\ldots.
\end{equation}
Note that the number operator has the same eigenfunctions as $\hat{H}$, but its eigenvalues are $k=0,1,2,\ldots$.

We would like to normalize the eigenfunctions of $\hat{H}$, i.e., find the coefficient $\gamma_n\in\mathbb{C}$ such that $\phi_n = \gamma_n \hat{a}^\dagger \phi_{n-1}$ satisfies $\langle \phi_n, \phi_n\rangle = 1$. If $n=1$, $\phi_0$ is already normalized and $\gamma_1=1$. If $n\geq 2$, let's assume that $\gamma_{n-1}$ is already known, such that $\phi_{n-1} = \gamma_{n-1}\hat{a}^\dagger \phi_{n-2}$ has unit norm. Then, from \eqref{eq_commute} and \eqref{eq_number-op},
\begin{multline*}
\langle \phi_n, \phi_n \rangle = \left|\gamma_n \right|^2 
\langle \hat{a}^\dagger \phi_{n-1}, \hat{a}^\dagger \phi_{n-1}\rangle =
\left|\gamma_n \right|^2\langle \phi_{n-1}, \hat{a}\hat{a}^\dagger \phi_{n-1}\rangle 
=\left|\gamma_n\right|^2\langle \phi_{n-1}, \left(\hat{a}^\dagger \hat{a}+\hat{I}\right) \phi_{n-1}\rangle \\
=\left|\gamma_n \right|^2\langle \phi_{n-1}, \left(n-1 + 1\right) \phi_{n-1}\rangle 
= n \left|\gamma_n \right|^2\langle \phi_{n-1},  \phi_{n-1}\rangle 
= n \left|\gamma_n \right|^2.
\end{multline*}
We have $\langle \phi_n, \phi_n \rangle = 1$ if $\gamma_n = 1/\sqrt{n}$. Thus, the normalized eigenfunctions satisfy
\begin{equation}
\phi_n = \frac{1}{\sqrt{n}}\hat{a}^\dagger \phi_{n-1} = \frac{1}{\sqrt{n(n-1)}}\left(\hat{a}^\dagger\right)^2 \phi_{n-2}=\ldots =
\frac{1}{\sqrt{n!}}\left(\hat{a}^\dagger\right)^n \phi_0,\quad n=1,2,3,\ldots.    
\end{equation}
In the quantum physics literature the normalized eigenfunction $\phi_n$ is often called the $n$-th eigenstate and is denoted by $|n\rangle$. 

Since $\hat{a}^\dagger \phi_{n-1} = \sqrt{n}\phi_n$, the relation \eqref{eq_commute} gives
\[
\hat{a} \sqrt{n}\phi_n = \hat{a}\hat{a}^\dagger \phi_{n-1} = \left(\hat{n} + \hat{I}\right)\phi_{n-1} = n\phi_{n-1}.
\]
In summary, the relations
\begin{equation}
    \hat{a} \phi_{n} = \sqrt{n} \phi_{n-1},\quad 
    \hat{a}^\dagger \phi_{n} = \sqrt{n+1}\phi_{n+1},
\end{equation}
motivate the names for $\hat{a}$ and $\hat{a}^\dagger$, i.e., applying the lowering operator $\hat{a}$ to the $n$-th eigenstate results in the $(n-1)$-th eigenstate, scaled by $\sqrt{n}$. Similarily, applying the raising operator $\hat{a}^\dagger$ to the $n$-th eigenstate results in the $(n+1)$-th eigenstate, scaled by $\sqrt{n+1}$.

\subsection{Heisenberg matrix formalism}
In the following we will assume that the wave function $\Psi$ belongs to a Hilbert space $\cal{V}$ with scalar product $\langle \cdot, \cdot \rangle$. Furthermore, assume that the Hamiltonian operator $\hat{H}$ is a linear compact self-adjoint operator on $\cal{V}$. Then, all eigenvalues $\lambda_j$ of $\hat{H}$ are real and the eigenfunctions form an orthonormal basis of $\cal{V}$. In many applications, the total Hamiltonian consists of a time-independent system Hamiltonian and a time-dependent control Hamiltonian,
\[
\hat{H} = \hat{H}_{sys} + \hat{H}_{ctrl}.
\]
For example, $\hat{H}_{sys}$ could be the quantum harmonic oscillator described in the previous section. While it may not be possible to compute the eigenfunctions of the total Hamiltonian, we can still expand the wave function in the eigenfunctions of the system Hamiltonian. 

In the following, we denote a set of orthonormal basis functions (not necessarily eigenfunctions) that span $\cal{V}$ by $B=\{\phi_k\}_{k=0}^\infty$. This means that
\begin{equation}\label{eq_orthonormal}
\langle \phi_n, \phi_m \rangle = \delta_{nm}, 
\end{equation}
and any function $u(\xb,t)\in{\cal V}$ can be expanded in the orthonormal basis,
\[
u(\xb,t) = \sum_{k=0}^\infty \tilde{u}_k(t) \phi_k(\xb),\quad \tilde{u}_j(t) = \langle \phi_j, u(\cdot, t) \rangle.
\]
In this way, any function $u(\xb,t)\in{\cal V}$ can be represented by the set of scalar coefficients $\{\tilde{u}_j(t)\}_{j=0}^\infty$, which we can collect in an (infinite-dimensional) vector $\ub = (\tilde{u}_0, \tilde{u}_1, \ldots)^T$.

We can expand the time-dependent wave function in terms of the elements of $B$,
\begin{equation}\label{eq_wave-expansion}
 \Psi(\xb,t) = \sum_{k=0}^\infty \psi_k(t) \phi_k(\xb).
\end{equation}
Inserting this expression into the Schr\"odinger equation~\eqref{eq_hamiltonian-abstract} gives
\[
i \hbar  \sum_{k=0}^\infty \frac{d \psi_k}{d t}(t) \phi_k(\xb)  = 
\sum_{k=0}^\infty \psi_k(t) \hat{H} \phi_k(\xb).
\]
By forming the scalar product between $\phi_j(x)$ and the above equation, we arrive at 
\[
i \hbar \frac{d \psi_j}{d t}(t) = \sum_{k=0}^\infty H_{jk} \psi_k,\quad j=0,1,2,\ldots,
\]
where we used the orthonormality relation~\eqref{eq_orthonormal}. This equation can be written as an infinite-dimensional system of ordinary differential equations (ODEs),
\begin{equation}\label{eq_schrodinger-inf-ode}
 \frac{d \psib}{d t} = - \frac{i}{\hbar} H \psib, \quad
\psib = \begin{bmatrix}
  \psi_0\\
  \psi_1\\
  \psi_2\\
  \vdots
\end{bmatrix},
\end{equation}
where the elements of the (infinite dimensional) Hamiltonian matrix $H$ are defined by
\begin{equation}
H_{jk} = \langle \phi_j, \hat{H}\phi_k \rangle,\quad j,k = 0, 1, 2, \ldots.
\end{equation}
Because the Hamiltonian operator $\hat{H}$ is self-adjoint, the Hamiltonian matrix $H$ must be Hermitian. It is common to scale the Hamiltonian matrix such that the factor $1/\hbar$ cancels out. 

In order to numerically solve the ODE formulation of Schr\"odinger's equation \eqref{eq_schrodinger-inf-ode}, it is necessary to truncate the series expansion \eqref{eq_wave-expansion} after some finite natural number $d$, resulting in a finite-dimensional system of ODEs. This representation of Schr\"odinger's equation will be used in the remainder of this document.

\section{Postulates of quantum physics}
\subsection{Postulate 1: State space}\label{subsec:postulate1}
From Lidar's lecture notes~\cite{lidar2020lecture}: The state of every closed (isolated) quantum system can be described by a
state vector that belongs to a Hilbert space $\cal H$. The Hilbert space is defined by
\begin{equation}\label{eq_hilbert}
  {\cal H} = \{\vb \in {\mathbb C}^d,\  \vb = \begin{bmatrix} v_0\\ v_1 \\ \vdots \\ v_{d-1} \end{bmatrix} |
  v_j \in {\mathbb C} \},
\end{equation}
where the scalar product and norm of the Hilbert space are defined by \eqref{eq_sp+norm}.

In addition, a quantum mechanical state must be normalized to unit length. 

From Wikipedia (Mathematical formulation of quantum mechanics): ``A quantum mechanical state is a ray in a projective Hilbert space, not a vector. Many textbooks fail to make this distinction, which could be partly a result of the fact that the Schrödinger equation itself involves Hilbert-space ``vectors", with the result that the imprecise use of ``state vector" rather than ray is very difficult to avoid."
  
\subsection{Postulate 2: Composite systems}
Given two quantum systems with respective Hilbert spaces ${\cal H}_1$ and ${\cal H}_2$, the state
space of the composite quantum system is ${\cal H} = {\cal H}_1 \otimes {\cal H}_2$.

\subsection{Postulate 3: Evolution}\label{subsec:postulate3}
The evolution of a closed quantum system is described by some unitary operator $U(t)$ such that the
state vector $\psib(t)$ evolves according to
\begin{equation}\label{eq_unitary}
\psib(t) = U(t) \psib(0), \quad t\geq 0.
\end{equation}
Equivalently, the state of the quantum system satisfies Schr\"odinger's equation,
\begin{equation}\label{eq_schrodinger}
\frac{\partial \psib}{\partial t} = -\frac{i}{\hbar} H(t) \psib,
\end{equation}
with $H=H^\dagger$ being a Hermitian operator known as the Hamiltonian. Here $\hbar$ is the reduced
Planck constant. As is mentioned above, it is common to scale the Hamiltonian such that the factor $1/\hbar$ cancels out.

\subsection{Postulate 4: Measurements} \label{subsec:postulate4}
Quantum measurements are described by a set of measurement operators $\{ M_k\}_{k=0}^{d-1}$ acting
on the state space of the system being measured. The index $k$ refers to the measurement outcomes
that may occur.

Given a quantum system that is in the state $\psib\in{\cal H}$ immediately before the
measurement, the probability that the measurement will result in outcome $k$ is
\begin{equation}\label{eq_m-prob}
  p_k = \| M_k \psib \|^2 = \langle \psib, M_k^\dagger M_k \psib \rangle.
\end{equation}
If outcome $k$ was observed, the state of the system is transformed according to
\begin{equation}\label{eq_m-trans}
  \psib \rightarrow \frac{M_k \psib}{\sqrt{p_k}} \equiv \psib_k.
\end{equation}
The state transformation is postulated to occur instantaneously after measurement.

Because the probabilities for measuring a state $\psib$ in one of the $d$ outcomes must sum to one,
\[
  1 = \sum_{k=0}^{d-1} p_k = \sum_{k=0}^{d-1} \langle \psib, M_k^\dagger M_k \psib \rangle =
  \langle \psib, \left(\sum_{k=0}^{d-1} M_k^\dagger M_k\right) \psib \rangle.
\]
Because $\psib$ is arbitrary, the measurement operators must satisfy the constraint
\begin{equation}\label{eq_m-constraint}
  \sum_{k=0}^{d-1} M_k^\dagger M_k = I. 
\end{equation}

\section{Density operator definition}

Consider the case where we don't know what state the system is in, but only that it comes from a mixture of states $\psib_i$ with respective probability $q_i$, for $i\in\{0,1,\dots,d-1\}$. This is called a pure state ensemble, $\{q_i, \psib_i\}_{i=0}^{d-1}$.

The probability density operator is used to characterize quantum systems whose state is not
completely known.

Let $\{ M_k\}_{k=0}^{d-1}$ be a set of measurement operators. Before measurement, assume the system
is in the state $\psib_i\in{\cal H}$. The measurement transforms the state according to
\[
\psib_i \rightarrow \frac{M_k \psib_i}{\sqrt{p_{k | i}}} = \psib_i^k,\quad \mbox{with probability}\quad
p_{k | i} = \psib_i^\dag M_k^\dag M_k \psib_i.
\]
Next consider the case where we don't know what state the system is in, but only that it comes from
a pure state ensemble, $\{q_i, \psib_i\}_{i=0}^{d-1}$. Then, the probability of obtaining outcome
$k$ by the measurement is
\begin{align*}
p_k = \sum_i q_i p_{k | i} = \sum_i  q_i \psib_i^\dag M_k^\dag M_k \psib_i =
\sum_i  q_i \mbox{Tr}(M_k^\dag M_k \psib_i \psib_i^\dag )
 = \mbox{Tr}\left(M_k^\dag M_k \sum_i  q_i \psib_i \psib_i^\dag \right) =  \mbox{Tr}\left(M_k^\dag M_k \rho \right),
\end{align*}
because $\ab^\dag \bb = \mbox{Tr}(\bb \ab^\dag)$ for all vectors $\ab$ and $\bb$ of equal
dimension. Here, we have introduced the density operator $\rho$ of the quantum system,
\begin{equation}\label{eq_densityOp}
 \rho = \sum_{i=0}^{d-1} q_i \psib_i \psib_i^\dag \in {\mathbb C}^{d\times d},\quad q_i \geq 0,\quad
\sum_{i=0}^{d-1} q_i = 1.   
\end{equation}

From the definition follows immediately that the density matrix is Hermitian (self-adjoint),
\begin{equation}
\rho^\dag = \sum_{i=0}^{d-1} q_i \psib_i \psib_i^\dag = \rho.
\end{equation}
The density operator is also positive semi-definite, because for all vectors $\vb\in{\mathbb C}^d$,
\begin{equation}
\vb^\dagger \rho \vb = \sum_{i=0}^{d-1} q_i \vb^\dagger \psib_i \psib_i^\dag \vb = \sum_{i=0}^{d-1}
q_i |\vb^\dagger \psib_i|^2 \geq 0,
\end{equation}
since the probabilities $q_i$ are real non-negative numbers.

Letting $\eb_k$ denotes the $k^{th}$ canonical unit vector, the $k^{th}$ diagonal element of $\rho$ equals $\eb_k^\dagger \rho \eb_k$ so that the trace of a density matrix satisfies
\begin{equation}
\mbox{Tr}(\rho) = \sum_{k=0}^{d-1}\sum_{i=0}^{d-1} q_i \eb_k^\dag \psib_i \psib_i^\dag \eb_k =
\sum_{i=0}^{d-1} q_i \sum_{k=0}^{d-1} |\eb_k^\dag \psib_i|^2 = \sum_{i=0}^{d-1} q_i \|  \psib_i \|^2= 1,
\end{equation}
because $\|\psib_i\|=1$ and the probabilities $q_i$ sum to one. 

Because $\rho$ is positive semi-definite, all of its eigenvalues must be non-negative. In addition,
since Tr$(\rho)$ equals the sum of the eigenvalues of $\rho$, the density matrix must have at least one
eigenvalue that is positive.

\section{The postulates of quantum physics for density matrices}\label{sec_dens-post}

\subsection{Postulate 1: State space}
Density matrices belong to the Hilbert-Schmidt space of linear operators $\rho\in{\mathbb
  C}^{d\times d}$ with Tr$(\rho)=1$, $\rho^\dagger = \rho$ and $\rho \geq 0$. This function space is
endowed with the inner product $\langle A, B\rangle_{HS} = \mbox{Tr$(A^\dagger B)$}$ for any two
operators $A$ and $B$ acting on the same Hilbert-Schmidt space. The inner product defines the length of a density matrix by
\[
\| \rho\| = \sqrt{\langle \rho, \rho \rangle_{HS}} = \sqrt{P}, \quad P = \mbox{Tr}(\rho^\dagger
\rho) = \mbox{Tr}(\rho^2).
\]
Here, the quantity $P$ is called the purity of the state and is bounded by $1/d \leq P \leq
1$. The state is pure if $P=1$ and mixed if $P<1$.

\subsection{Postulate 2: Composite systems}
The state space for a density matrix describing a composite quantum system belongs to the
Hilbert-Schmidt space formed by Kronecker products between the Hilbert-Schmidt spaces of the
subsystems.

\subsection{Postulate 3: Evolution} Density matrices evolve according to the Liouville-von Neumann
equation,
\begin{equation}
  \dot{\rho} = -i[H,\rho],
\end{equation}
under a Hamiltonian operator $H=H^\dagger$, or equivalently as $\rho(t) = U(t)\rho(0)U^\dagger(t)$, where
$U(t)$ is a unitary operator.

\subsection{Postulate 4:  Measurement}
A general measurement defined by the set of operators $\{M_k\}_{k=0}^{d-1}$ results in outcome $k$
with probability $p_k = \mbox{Tr}(M_k^\dagger M_k \rho)$. The density matrix is transformed
according to
\[
\rho \mapsto \frac{M_k \rho M_k^\dagger}{p_k}.
\]

\newpage
\section{Closed quantum systems}\label{sec_closed}
In this section we are looking for definitions in terms of the state vector, usually denoted by $\psib$.

\subsection{Closed quantum system definition}
A closed quantum system is isolated from the surrounding environment and preserves certain invariants such as energy over time. A closed quantum system is modeled by Schr\"{o}dinger's equation (\ref{eq_schrodinger}) as required by Postulate 3 in Section~\ref{subsec:postulate3}. Conservation of the norm of $\psib$ is  manifested through the equality 
\[
\frac{d \psib^\dagger \psib}{dt} = 0,
\]        
which follows from Schr\"{o}dinger's equation together with the Hermiticity of $H$.

\subsection{Normalization of a state vector}
As the norm of $\psib$ is conserved in time it is sufficient to require that the initial state vector has norm one. If this is true the state vector will also satisfy Postulate 1 in Section~\ref{subsec:postulate1}, i.e. it belongs to the Hilbert space $\cal{H}$. The condition that $\psib^\dagger \psib = 1$ is also referred to as the normalization condition.

\subsection{What is a qubit?}
A qubit or quantum bit is the basic unit of quantum information, just as a bit is the basic unit of information in classical information theory.   

When the dimension of the Hilbert space is 2, the state vector $\psib$ represents the state of the qubit. Then, unlike a classical bit which only has two states,  we may have a superposition of states. That is, all states are of the form
\[
\psib = \alpha \left[ \begin{array}{cc}
     1  \\
     0 
\end{array} \right]
+ \beta \left[ \begin{array}{cc}
     0  \\
     1 
\end{array} \right],\quad \alpha,\beta \in \mathbb{C},
\]
with the normalization condition $|\alpha|^2 + |\beta|^2 = 1$. The above state is expressed as a linear combination of the  computational basis vectors. In  Dirac notation we would write
\[
\psib = \alpha | 0 \rangle + \beta | 1\rangle.
\]
\subsection{What does $|01\rangle$ mean?}
In Dirac notation, $|01\rangle$ is a basis vector in a composite system consisting of two sub-systems, $A$ and $B$. It is short hand for
\[
|01\rangle = |0\rangle_A \otimes |1\rangle_B = \eb^A_0 \otimes \eb^B_1.
\]
where $\eb^{A,B}_j$ denote the canonical basis vectors in systems $A$ and $B$, which can be of different dimensions. If both sub-systems have dimension two (qubits), the basis vector $|01\rangle$ is a column vector of dimension four,
\[
|01\rangle = 
\begin{bmatrix} 1 \\ 0 \end{bmatrix}\otimes
\begin{bmatrix} 0 \\ 1 \end{bmatrix} =
\begin{bmatrix} 0 \\ 1 \\ 0 \\ 0\end{bmatrix}.
\]

\subsection{Entangled state}
An entangled state is a state that cannot be written as a product of states. The canonical example is the pure two qubit Bell state
\[
\psib = \frac{1}{\sqrt{2}} (|00\rangle + |11\rangle) =
\frac{1}{\sqrt{2}}
 \begin{bmatrix}
      1 \\ 0 \\ 0 \\ 0
    \end{bmatrix}
+
\frac{1}{\sqrt{2}}
\begin{bmatrix}
      0 \\ 0 \\ 0 \\ 1
    \end{bmatrix}
\]
\subsection{Bell state}
The above is the first of four Bell states, which are maximally entangled (see Section~\ref{sec_max-entangled}) states of a two qubit system. The remaining three are  
\[
\frac{1}{\sqrt{2}} (|00\rangle - |11\rangle) =
\frac{1}{\sqrt{2}}
 \begin{bmatrix}
      1 \\ 0 \\ 0 \\ 0
    \end{bmatrix}
-
\frac{1}{\sqrt{2}}
\begin{bmatrix}
      0 \\ 0 \\ 0 \\ 1
    \end{bmatrix},
\quad 
\frac{1}{\sqrt{2}} (|01\rangle + |10\rangle) =
\frac{1}{\sqrt{2}} 
 \begin{bmatrix}
      0 \\ 1 \\ 0 \\ 0
    \end{bmatrix}
+
 \frac{1}{\sqrt{2}}
 \begin{bmatrix}
      0 \\ 0 \\ 1 \\ 0
    \end{bmatrix},
\]
and
\[
\frac{1}{\sqrt{2}} (|01\rangle - |10\rangle) =
\frac{1}{\sqrt{2}}
 \begin{bmatrix}
      0 \\ 1 \\ 0 \\ 0
    \end{bmatrix}
-
 \frac{1}{\sqrt{2}}\begin{bmatrix}
      0 \\ 0 \\ 1 \\ 0
    \end{bmatrix}.
\]

\subsection{Bloch sphere}
The Bloch sphere is a geometrical representation of the pure state space of a two-level quantum mechanical system (qubit), named after the physicist Felix Bloch. Quantum mechanics is mathematically formulated in a Hilbert space or a projective Hilbert space. For a two-dimensional Hilbert space, the space of all pure states is the complex projective line $\mathbb{CP}^1$. This is the Bloch sphere, also known to mathematicians as the Riemann sphere. 

The Bloch sphere is a unit 2-sphere, with antipodal points corresponding to a pair of mutually orthogonal state vectors. The north and south poles of the Bloch sphere are typically chosen to correspond to the standard basis vectors $|0\rangle$ and $|1\rangle$, respectively, which in turn might correspond e.g. to the spin-up and spin-down states of an electron. This choice is arbitrary, however. The points on the surface of the sphere correspond to the pure states of the system, whereas the interior points correspond to the mixed states. The Bloch sphere may be generalized to an $n$-level quantum system, but then the visualization is less useful. 

Consider a two-level quantum system in the pure state $\psib$, which we may write as $\psib = \alpha [1,0]^T + \beta [0,1]^T$, where $\alpha, \, \beta \in \C$ or, alternatively, $\alpha = a + i b,$ $\beta = c + i d$ where $a, \, b, \, c,\,d \in \R$ and $i$ is the imaginary unit. The requirement that $\|\psib\| = 1$ implies that there are only three degrees of freedom in choosing $a, \, b, \, c,\,d \in \R$ and we may thus visually represent the state in a three-dimensional space.  Since only the relative phase between the two coefficients $\alpha, \, \beta$ is important and $\|\psib\| =1$, we may, without loss of generality, assume that  $\boldsymbol{\alpha}$ {\bf is real and non-negative} so that we can instead write the state as $\psib = \cos(\theta/2) [1,0]^T + e^{i\varphi} [0,1]^T$ with $0 \le \theta \le \pi$ and $0 \le \varphi < 2\pi$. Letting $\vec{v} = (\sin\theta \cos\varphi , \, \sin\theta \sin \varphi, \, \cos\theta)$, we note that this (unit) vector can be interpreted as being in spherical coordinates and so is on the surface of the unit ball in $\R^3$ as shown below.
\begin{figure}[H]
\graphicspath{{Plots/}}
  \centering
    \includegraphics[width=0.3\textwidth]{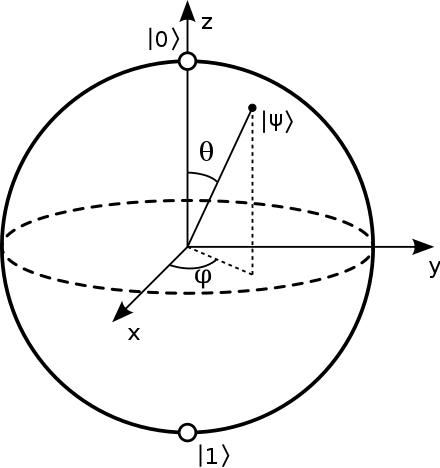}
    \caption[]%
    {{\small The usual Bloch sphere using spherical coordinates to describe a state vector, $\psib$.}}
    \label{fig:bloch}
\end{figure}

In general, we can consider the density matrix $\rho \in \C^{2\times 2}$ and recall that density matrices are Hermitian and have unit trace. Any such $\rho$ can be written as 
  \begin{align*}
    \rho = \frac{1}{2} \left(I + \sum_i v_i \sigma_i \right) = \frac{1}{2} \left(I + \vec{v} \cdot \vec{\sigma} \right),
  \end{align*}
where $\vec{v} \in \R^3$ and $\sigma_i$ are the Hermitian, traceless Pauli matrices,
  \begin{align*}
    \sigma_1 = \sigma_x = \begin{pmatrix}0 & 1\\ 1 & 0\end{pmatrix},\quad
    \sigma_2 = \sigma_y = \begin{pmatrix}0 & -i\\ i & 0\end{pmatrix},\quad
    \sigma_3 = \sigma_z = \begin{pmatrix}1 & 0\\ 0 & -1\end{pmatrix}.
  \end{align*}
By the above construction we guarantee unit trace, but to enforce positivity of the density matrix one can show that the eigenvalues of $\rho$ are 
  \begin{align*}
    \lambda_{\pm} = \frac{1}{2} \left(1 \pm \|\vec{v}\|\right),
  \end{align*}
which requires $\|\vec{v}\| \le 1$. We call $\vec{v}$ \textit{the Bloch vector}. Note that for a pure quantum state we require $\rho^2 = \rho$, and we have that $\text{Tr}\rho^2 = (1 + \|\vec{v}\|^2)/2$ so that any unit Bloch vector is a pure state, otherwise we have a mixed state.

\subsection{Unitary evolution}\label{sec_unitary_evolution}
For a quantum system we say that the state $\psib$ (or alternatively the density matrix $\rho$) has \textit{unitary evolution} if the solution is of the form $\psib(t) = U(t) \psib_0$ where $\psib_0 \in \C^d$ (or $\rho(t) = U(t) \rho_0 U(t)^\dagger$ where $\rho_0 \in \C^{d\times d}$) and $U$ is a unitary operator (i.e. $U(t)^\dagger U(t) = U(t) U(t)^\dagger = I$). By definition, it follows that 
\begin{align*}
  \|\psib (t) \|^{2} = \|U(t) \psib_0\|^{2} = \left(U(t) \psib_0\right)^\dagger U(t) \psib_0 = \psib_0^\dagger U^\dagger(t)U(t) \psib_0 = \psib_0^\dagger \psib_0 = \|\psib_0\|^{2},
\end{align*}
with an analagous result for density matrices. It follows that unitary evolution necessarily preserves norm.

\subsection{Lowering matrix}
The lowering matrix `$a$' represents the action of the lowering operator $\hat{a}$ on the eigenstate expansion of the wave function \eqref{eq_wave-expansion}. Let $\Psi$ and $\Gamma$ be expanded in the eigenstates $\phi_k$ with coefficients $\psi_k$ and $\gamma_k$, respectively. Then
\begin{align}
\Gamma = \hat{a}\Psi,\quad \Leftrightarrow\quad
 \gamma_j = \sum_{k=0}^d a_{jk}\psi_k,\quad 
    a := \begin{pmatrix} 0  & \sqrt{1} & \\ 
                            &  0       & \sqrt{2}  \\
                            &       & \ddots & \ddots  \\
                            &       & &  0    &  \sqrt{d-1} \\
                            &       &        &  &0  
        \end{pmatrix} \in \mathbb C^{d\times d}.
\end{align}
Its action on $|n\rangle = \eb_n = (0,\dots,1,\dots,0)^T$ for $n=1,\dots,d-1$ is
\begin{align}
    a|n\rangle = \sqrt{n} |n-1\rangle,
\end{align}
and for $n=0$, we get $a|0\rangle =0$.

\subsection{Raising operator}
The raising operator is the conjugate transpose of the lowering operator:
\begin{align}
    a^\dagger = \begin{pmatrix} 0  &  \\
            \sqrt{1} &  0    \\
            &   \ddots & \ddots  \\
            &   &  \sqrt{d-1} & 0 \\
        \end{pmatrix} \in \mathbb C^{d\times d}.
\end{align}
Its action on $|n\rangle = \eb_n= (0,\dots,1,\dots,0)^T$ for $n=0,\dots, d-2$ is
\begin{align}
    a^\dagger|n\rangle = \sqrt{n+1} |n+1\rangle,
\end{align}
and $a^\dagger|d-1\rangle = 0$ (this is an consequence of truncation to finite dimension).
\subsection{Number operator}
 The number operator is defined as
 \begin{align}
    N := a^\dagger a = \begin{bmatrix}
      0 \\ & 1 \\ & & \ddots \\ & & & d-1
    \end{bmatrix}.
 \end{align}
Its action on $|n\rangle = \eb_n= (0,\dots,1,\dots,0)^T$ is
\begin{align}
    N|n\rangle = n |n\rangle.
\end{align}
    
\subsection{Measurement outcome}\label{Measurement}
(The following is a restatement of Postulate 4 in Section~\ref{subsec:postulate4}.) 

A measure in quantum mechanics consists of a set of measurement operators $\{M_k\}_{k=0}^{d-1}$. The index $k$ refers to the possible outcomes of the measurement. Consider a quantum system that is in the state $\psib\in{\cal H}$ immediately before the measurement. The probability that the measurement will result in outcome $k$ is
\begin{equation}
  p_k = \| M_k \psib \|^2 = \langle \psib, M_k^\dagger M_k \psib \rangle.
\end{equation}

Because the probabilities for measuring a state $\psib$ in one of the $d$ outcomes must sum to one,
\[
  1 = \sum_{k=0}^{d-1} p_k = \sum_{k=0}^{d-1} \langle \psib, M_k^\dagger M_k \psib \rangle =
  \langle \psib, \left(\sum_{k=0}^{d-1} M_k^\dagger M_k\right) \psib \rangle.
\]
Since $\psib$ is arbitrary, the measurement operators must satisfy the constraint
\begin{equation}
  \sum_{k=0}^{d-1} M_k^\dagger M_k = I. 
\end{equation}

For example, for an observable $A$ (i.e. a Hermitian operator) associated to a physical quantity, the measurement outcomes are the eigenvalues of $A$. The measurement operators are $P_a = |a\rangle \langle a|$ where $|a\rangle$ are the (orthonormal) eigenvectors of $A$.

\subsection{Measurement transformation}
(The following is a restatement of Postulate 4 in Section~\ref{subsec:postulate4}.) 

Quantum measurements are described by a set of measurement operators $\{ M_k\}_{k=0}^{d-1}$ acting on the state space of the system being measured. The index $k$ refers to the measurement outcomes that may occur.

If outcome $k$ was observed, the state of the system is transformed according to
\begin{equation}
  \psib \mapsto \frac{M_k \psib}{\sqrt{p_k}} \equiv \psib_k.
\end{equation}
The state transformation is postulated to occur instantaneously after measurement.

\subsection{Positive operator valued measure}\label{subsec:POVm}
Based on a general set of measurement operators $\{M_k\}_{k=1}^d$, the set of Hermitian operators $F_k = M_k^\dagger M_k$, $k=1,2,\ldots,d$, form a positive operator valued measure (POVM). The POVM elements $F_{k}$ must be normalized such that
\begin{equation}
    \sum_{k=1}^d F_k = I,\quad F_k = M_k^\dagger M_k.
\end{equation}
The probability of obtaining measurement outcome $k$ when applied to a state vector $\psib$ follows from Postulate 4 in Section~\ref{subsec:postulate4},
\begin{equation}
  p_k = \langle \psib, M_k^\dagger M_k \psib\rangle = \langle \psib, F_k \psib\rangle =
  \operatorname {Tr} (\psib \psib^\dagger F_{k}),  
\end{equation}
where $\operatorname {Tr}$ is the trace operator. 


\subsection{Observable}
Every physically measurable quantity is associated with an observable, i.e, a Hermitian operator $A$. Since $A$ is Hermitian it has a spectral decomposition,
\begin{equation}
 A =\sum_k \lambda_k \ab_k \ab_k^\dagger, \quad \ab_k^\dagger \ab_k = 1,
\end{equation}
where $\ab_k$ is the eigenvector corresponding to the eigenvalue $\lambda_k$. All eigenvalues $\lambda_k \in \mathbb{R}$ since $A$ is Hermitian and 
are the possible outcomes of the measurement. The probability of measuring the outcome $\lambda_k$ is given by
\begin{align}\label{eq_probabilityLambdak}
    p_k = \psib^\dagger P_k^\dagger P_k\psib = |\psib^\dagger \ab_k |^2,
\end{align}
where $P_k = \ab_k \ab_k^\dagger$
(compare Section \ref{subsec:POVm}, \ref{subsec:projectivemeasure}).

\subsection{Projective measurement}\label{subsec:projectivemeasure}
Projective measurements are a special case of generalized measurements, in which the measurement operators $M_k$ are Hermitian operators called projectors, denoted by $P_k=\ab_k \ab_k^\dagger$. That is, $M_k = P_k$, where $P_k P_l = \delta_{k,l}P_k$ and $P_k^{\dagger}=P_k$. In particular, $P^2_k = P_k$. Using this we can see that the probability outcome of $k$ is $p_k = \langle \psi | M^{\dagger}_k M_k | \psi \rangle = \langle\psi | P_k | \psi \rangle$.

\subsection{Expectation of an observable}
Let $A$ be an observable 
\begin{equation}\label{eq_observable2}
 A =\sum_k \lambda_k \ab_k \ab_k^\dagger, \quad P_k = \ab_k \ab_k^\dagger.
\end{equation}
Since we obtain $\lambda_k$ with probability $p_k$, we naturally define an expectation of this observable in the state $| \YC{\psi} \rangle$ as
\begin{equation}\label{eqn::Expectation}
    \langle A \rangle_{\psi} := \sum_{k} \lambda_k p_k = \mbox{Tr}(A \psib \psib^\dagger) =  \psib^\dagger A \psib.
\end{equation}

\subsection{Standard deviation of an observable}
For a random variable $X$ with mean $\mu$ we can define the standard deviation as the square root of the variance of $X$, i.e. $\mathbb{E}[(X-\mu)^2]^{1/2}$, where $\mathbb{E}$ denotes the expected value of a random variable. Using the definition of the expectation of an observable \eqref{eqn::Expectation}, we can define the standard deviation of an observable in the state $\psib$ as 
\begin{align}\label{eqn::StandardDeviation}
    \Delta A = \sqrt{\bigg \langle \big (A - \langle A \rangle_\psi I\big)^2 \bigg\rangle_\psi}
    = \sqrt{\psib^\dag (A - \langle A \rangle_\psi I)^2 \psib}
    & = \sqrt{\psib^\dag (A - \psib^\dag A \psib I)^2 \psib} \nonumber \\
    & = \sqrt{\psib^\dag A^2 \psib - (\psib^\dag A \psib)^2 I} \nonumber \\ 
    & = \sqrt{\sum_k \lambda_k^2 p_k - \left(\sum_k \lambda_k p_k\right)^2}.
\end{align}
Note that if $\psib$ is the j$^{th}$ eigenvector of the observable $A$ then $p_j = \delta_{jk}$ so that $\Delta A =0$.

\subsection{Global phase equivalence}
Two quantum states, $\psib$ and $\phib$, are said to be equivalent if they only differ by a global phase factor, i.e.,
\[
\phib = e^{i\theta} \psib,\quad \theta \in \mathbb{R}.
\]

\subsection{Ehrenfest's theorem}
Let $\psib$ satisfy Schr\"odinger's equation with the Hamiltonian $H$. Then the expectation of the observable operator $A$ evolves according to
\[
\frac{d}{dt}\langle A \rangle_\psi = \frac{1}{i\hbar} \langle [A,H]\rangle_\psi.
\]
\subsection{Rotating frame transformation}
Consider the Schr\"odinger equation in a laboratory frame of reference,
\begin{equation}\label{eq_schrodinger2}
\dot{\psib} = -i H(t) \psib,\quad 0\leq t \leq T,\quad \psib(0) = \psib_0.
\end{equation}
Here, $\psi(t) \in [0,T]\to {\mathbb C}^d$ is the state vector and $H(t) \in [0,T]\to {\mathbb C}^{d\times d}$ is the Hamiltonian matrix.

Consider the unitary transformation
\[
\psib(t) = R^{\dag}(t)v(t),\quad R^\dag R = I.
\]
We have
\begin{align*}
\dot{\psib} = \dot{R}^\dag v + R^\dag \dot{v},\quad
H\psib = H R^\dag v.
\end{align*}
Thus, \eqref{eq_schrodinger2} gives
\[
\dot{R}^\dag v + R^\dag \dot{v} = -i H R^\dag v,
\]
By using the identity $R \dot{R}^\dag = - \dot{R} R^\dag$ and reorganizing the terms,
\[
\dot{v} = -i R H R^\dag v + \dot{R} R^\dag v = -i\left( RHR^\dag + i \dot{R} R^\dag \right) v.
\]
Thus, the transformed problem becomes
\begin{equation}\label{eq_timedep_trans}
\dot{v} = -i \tilde{H}(t) v,\quad \tilde{H}(t) = R(t)H(t)R^\dag(t) + i \dot{R}(t) R^\dag(t).
\end{equation}

When the Hamiltonian is of the form,
\begin{equation}\label{eq_quantum-osc}
H(t) = \omega_a N + H_d + H_c(t),
\quad 
N = \begin{bmatrix}
  0 & & & & \\
  & 1 & & & \\
  && 2 && \\
  &&& \ddots & \\
  &&&& d-1
\end{bmatrix},
\end{equation}
the difference between consecutive diagonal elements in $\omega_a N$ is constant. This structure suggests the unitary transformation
\[
R(t) = \exp(i\omega_a N t),\quad \dot{R}R^\dag = i\omega_a N.
\]
The matrices $N$ and $R(t)$ are diagonal and therefore commute. The first term in the Hamiltonian \eqref{eq_quantum-osc} cancels \YC{with $i \dot{R}(t) R^\dag(t)$ in \eqref{eq_timedep_trans}} and the transformed Hamiltonian becomes
\begin{equation}\label{eq_rot_hamiltonian0}
  \tilde{H}(t) = R(t)(H_d + H_c(t))R^\dag(t).
\end{equation}

The rotating frame transformation can be generalized to cancel terms in Hamiltonians that are of the form
\[
\omega_a N \otimes I + \omega_b I \otimes N.
\]
In that case the unitary transformation becomes
\[
R(t) = \exp(i\omega_a N t)\otimes I + I \otimes \exp(i\omega_b N t).
\]
The construction can be generalized further. 

\subsection{Rotating wave approximation}\label{RWA}

We illustrate the rotating wave approximation (RWA) in the case when the control Hamiltonian is of the form,
\[
H_c(t) = f(t)(a + a^\dagger),\quad a = \begin{bmatrix}
0 & 1 & & & &\\
 & 0 & \sqrt{2} & & &\\
&  & 0 & \sqrt{3} & &\\
& &  & 0 & \sqrt{4} & \\
& &  &  & \ddots & \ddots\\
\end{bmatrix}.
\]
Here, $a^\dag$ is the Hermitian conjugate of the matrix $a$
and $f(t)$ is a real-valued function of time. The rotating frame transformation results in
\begin{equation}\label{eq_rot_hamiltonian}
\tilde{H}_c(t) = f(t) \left( R(t) a R^\dag(t) + R(t) a^\dag R^\dag(t) \right).
\end{equation}
We have
\begin{multline*}
  R a R^\dag =\\
\begin{bmatrix}
  1 & & & \\
  & e^{i\omega_a t} & & \\
  & &  e^{2i\omega_a t} & \\
  & & & \ddots
\end{bmatrix}
\begin{bmatrix}
0 & 1 & & &\\
 & 0 & \sqrt{2} & &\\
&  & 0 & \sqrt{3} &\\
& &  & \ddots & \ddots
\end{bmatrix}
\begin{bmatrix}
  1 & & & \\
  & e^{-i\omega_a t} & & \\
  & &  e^{-2i\omega_a t} & \\
  & & & \ddots
\end{bmatrix} = \\
\begin{bmatrix}
0 & e^{-i\omega_a t} & & &\\
 & 0 & \sqrt{2} e^{-i\omega_a t}& &\\
&  & 0 & \sqrt{3} e^{-i\omega_a t}  &\\
& &  & \ddots & \ddots
\end{bmatrix} =: e^{-i\omega_a t} a.
\end{multline*}
Taking the conjugate transpose gives $R a^\dag R^\dag = e^{i\omega_a t} a^\dag$. Thus,
\eqref{eq_rot_hamiltonian} becomes
\begin{equation}\label{eq_trans-hamiltonian}
\tilde{H}_c(t) =
  f(t) \left( e^{-i\omega_a t} a + e^{i\omega_a t} a^\dag \right).
\end{equation}

The RWA aims to absorb the highly oscillatory factors $\exp(\pm i\omega_a t)$ into
$f(t)$. Because the control function $f(t)$ is real-valued, this can only be done in an approximate fashion. We make the ansatz,
\begin{equation}
  f(t) = 2p(t) \cos(\omega_a t) - 2q(t) \sin(\omega_a t) = 
  \left(p + i q\right)\exp(i\omega_a t) + \left( p - i q \right)
  \exp(-i\omega_a t),
\end{equation}
where $p(t)$ and $q(t)$ are real-valued function. Thus,
\begin{align}
f(t)e^{-i\omega_a t} &= \left(p + iq\right) +  \left(p - iq\right)
e^{-2i\omega_a t},\\
f(t)e^{i\omega_a t} &= \left(p + i q\right)e^{2i\omega_a t} +  \left(p - iq\right).
\end{align}
The transformed Hamiltonian \eqref{eq_trans-hamiltonian} becomes
\begin{align*}
\widetilde{H}_c(t) &= \left(p + i q\right) a +  \left(p - i q\right) e^{-2i\omega_a t} a +
\left(p + i q\right)e^{2i\omega_a t} a^\dag +  \left(p - i q\right) a^\dag \\
&= p\left( a + a^\dag \right) + i q \left( a - a^\dag \right)
\\
&\quad +
\left(p - iq\right) e^{-2i\omega_a t} a  + \left(p + iq\right)e^{2i\omega_a t} a^\dag.
\end{align*}
The rotating frame approximation follows by ignoring the terms that oscillate with twice the frequency, resulting in the approximate control Hamiltonian
\begin{align}
\hat{H}_c(t) &=  p(t)\left( a + a^\dag \right) + i q(t) \left( a - a^\dag
\right).
\end{align}

\subsection{Rabi oscillation}
Consider Schr\"odinger's equation for a two-level system in a
rotating frame of reference. Because we are in the rotating frame, the drift Hamiltonian is
zero. Let the control Hamiltonian be constant in time,
\begin{equation}\label{eq_rabi-again}
\dot{\psib} = -i H_c \psib,\quad 0 \leq t \leq T,
\end{equation}
where
\begin{equation}\label{eq_ctrl-ham}
H_c = \Omega a + \bar{\Omega} a^\dag,\quad a=\begin{bmatrix} 0 & 1 \\ 0 & 0 \end{bmatrix},\quad
 \Omega\in{\mathbb C}.
\end{equation}
This problem is known as a Rabi oscillator and can be solved analytically through an eigenvector
decomposition of the Hamiltonian matrix.
To derive the solution
we start by noting that $H_c$ has the eigenvector decomposition
\[
H_c X = X \Lambda,\quad
X = \frac{1}{\sqrt{2}}\begin{bmatrix}
  1 & -\Omega/|\Omega| \\
  \bar{\Omega}/|\Omega| & 1
\end{bmatrix},\quad
\Lambda = \begin{bmatrix}
  |\Omega| & 0 \\
  0 & -|\Omega|
\end{bmatrix}.
\]
Because $X$ is independent of time and unitary, the variable substitution $\tilde{\psib} = X^\dag \psib$ leads to
a decoupled system that can be solved analytically,
\[
\dot{\tilde{\psib}} = -i \Lambda \tilde{\psib}, \quad \tilde{\psib}(t) =
\begin{bmatrix}
  A \exp(-i|\Omega| t)\\
  B \exp(i|\Omega| t)
\end{bmatrix}.
\]
Transforming back to the original variable gives
\[
\psib(t) = \frac{1}{\sqrt{2}}
\begin{bmatrix}
A \exp(-i|\Omega| t) - B\frac{ \Omega}{|\Omega|}\exp(i|\Omega| t) \\
A\frac{ \bar{\Omega}}{|\Omega|}\exp(-i |\Omega| t) + B \exp(i|\Omega| t)
\end{bmatrix}.
\]
To construct the solution operator matrix, we must consider two initial conditions. First,
$\psib^{(1)}(0) = [ 1, 0]^T$, which gives $A_1=1/\sqrt{2}$ and
$B_1=-\bar{\Omega}/(\sqrt{2}|\Omega|)$. Secondly, $\psib^{(2)}(0) = [ 0, 1 ]^T$, which gives
$A_2=\Omega/(\sqrt{2}|\Omega|)$ and $B_2=1/\sqrt{2}$. After some algebra,
\begin{equation}\label{eq_sol-op}
U(t) =
\begin{bmatrix}
  \cos(|\Omega| t) &   (\sin(\theta) - i\cos(\theta)) \sin(|\Omega| t)\\
  -(\sin(\theta) + i\cos(\theta)) \sin(|\Omega| t) &   \cos(|\Omega| t)
\end{bmatrix},
\end{equation}
where the phase angle $\theta$ satisfies
\[
\Omega = |\Omega|(\cos(\theta) + i \sin(\theta)).
\]
Thus, the general solution of \eqref{eq_rabi-again} satisfies $\psib(t) = U(t)\psib(0)$, where $U(t)$ is called the solution operator matrix.

\begin{figure}
\graphicspath{{Plots/}}
  \centering
  \includegraphics[width=0.7\textwidth]{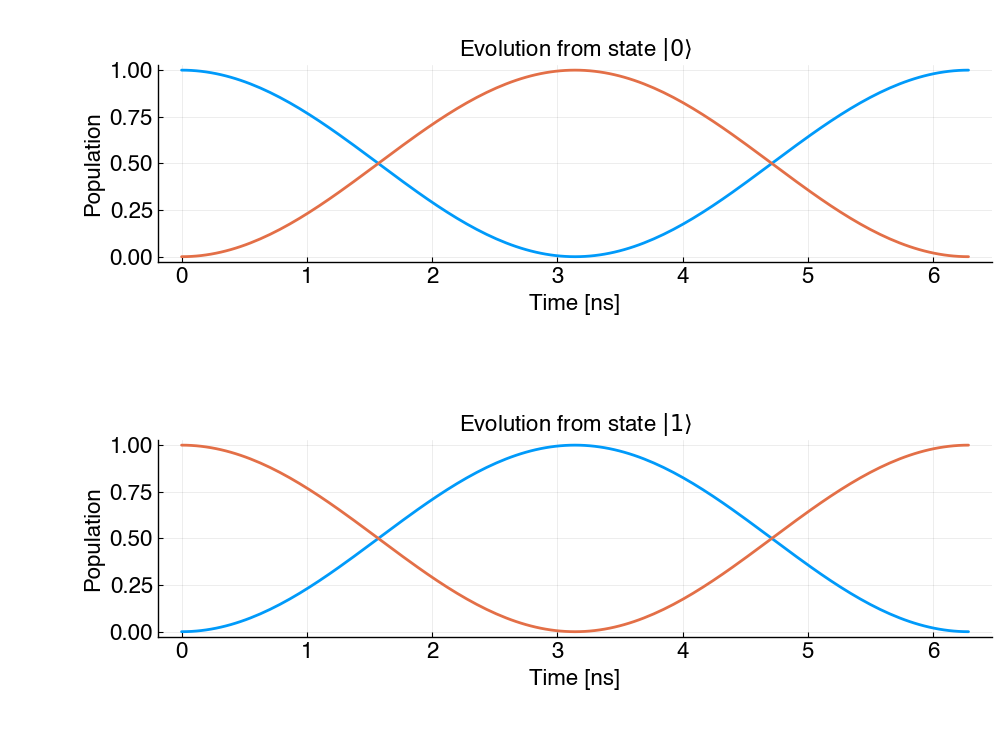}
  \caption{The evolution of the state probabilities during one period of a Rabi oscillation with pulse amplitude $|\Omega|=0.5$, showing $|\psi_0^{(j)}(t)|^2$ (blue) and $|\psi_1^{(j)}(t)|^2$ (orange), for the cases $j=0$ (top) and
  $j=1$ (bottom).}
  \label{fig:one-rabi}
\end{figure}

Let the components of the state vector be $\psib = [\psi_0, \psi_1]^T$. The measurement operators corresponding to the ground and excited states are
\[
M_0 = \begin{bmatrix} 1 & 0 \\ 0 & 0 \end{bmatrix},\quad
M_1 = \begin{bmatrix} 0 & 0 \\ 0 & 1 \end{bmatrix}.
\]
From Postulate 4 in Section~\ref{subsec:postulate4}, the probability of measuring $\psib(t)$ in the ground and excited states equal
$\| M_0 \psib(t)\|^2 = |\psi_0(t)|^2$ and $\| M_1 \psib(t) \|^2 = |\psi_1(t)|^2$, respectively. For example, starting from the ground state, $\| M_0 \psib(t)\|^2 = \cos^2(|\Omega| t)$ and $\| M_1 \psib(t)\|^2 = \sin^2(|\Omega|t)$. Thus, the probabilities oscillate harmonically in time with period
\[
\tau_p = \frac{\pi}{|\Omega|}.
\]
Note that the period of the oscillation is inversely proportional to the amplitude of the control signal.
Let the state vector satisfy the initial conditions $\psib^{(j)}(0)=\eb_j$ for $j=0,1$. The evolution of the probabilities $|\psi_0^{(j)}(t)|^2$ and $|\psi_1^{(j)}(t)|^2$ for $\theta = \pi/2$
and $|\Omega|=0.5$ are illustrated in Figure~\ref{fig:one-rabi}.

The solution operator matrix \eqref{eq_sol-op} can be used to construct analytical solutions of the quantum control problem.
For example, at $t=\tau_p/4$, it satisfies,
\begin{equation}\label{eq_rabi-uni}
U(\tau_p/4) = \frac{1}{\sqrt{2}}
\begin{bmatrix}
  1 &   (\sin(\theta) - i\cos(\theta)) \\
  -(\sin(\theta) + i\cos(\theta))  &  1
\end{bmatrix}.
\end{equation}
Note that the control Hamiltonian \eqref{eq_ctrl-ham} can be written
\[
H_c = p_0(a+a^\dag) + iq_0(a-a^\dag),\quad p_0 = \mbox{Re$\,(\Omega)$},\quad q_0 = \mbox{Im$\,(\Omega)$}
\]
Hence, the control problem for realizing the unitary transformation \eqref{eq_rabi-uni} with a
duration of $T=\pi/(4|\Omega|)$ has the analytical solution
\begin{equation}
p_0 = |\Omega|\cos(\theta),\quad q_0 = |\Omega|\sin(\theta).
\end{equation}

\subsection{$\pi$-pulse and $\pi/2$-pulse}
A $\pi$-pulse corresponds to one half of a Rabi oscillation. This pulse swaps the states $|0\rangle$ and $|1\rangle$. From the derivation in the previous section follows that the duration of a $\pi$-pulse with amplitude $|\Omega|$ is
\begin{align*}
    \tau_{\pi} = \frac{1}{2}\tau_p = \frac{\pi}{2|\Omega|}.
\end{align*}
Correspondingly, a $\pi/2$-pulse transforms the $|0\rangle$ and $|1\rangle$ states to an equal superposition of the $|0\rangle$ and $|1\rangle$ states. In terms of the Bloch sphere, a $\pi/2$ pulse moves states on either of the poles to the equator. The duration of a $\pi/2$-pulse with amplitude $|\Omega|$ is
\begin{align*}
    \tau_{\pi/2} = \frac{\pi}{4|\Omega|}.
\end{align*}

\subsection{Accuracy of the rotating wave approximation}
Let $\boldsymbol{u}$ denote the solution to the transformed Schr\"odinger equation from Section \ref{RWA}
\begin{align*}
  \frac{d}{dt}\boldsymbol{u} = -i\widetilde{H}(t) \boldsymbol{u},
\end{align*}
and analogously let $\boldsymbol{v}$ denote the solution to the rotating wave approximation attained by ignoring the
more oscillatory terms of $\tilde{H}$
\begin{align*}
  \frac{d}{dt}\boldsymbol{v} = -i\left[-\frac{\xi_a}{2} \left((a^\dag a)^2 - a^\dag a \right) + p(t)\left( a + a^\dag \right) + i q(t) \left( a - a^\dag \right)\right] \boldsymbol{v}.
\end{align*}
We can define the error $\boldsymbol{e} = \boldsymbol{u} - \boldsymbol{v}$ and note that it satisfies a related (inhomogeneous) ODE which we can analytically solve. If we let $g(t) = p(t) + iq(t)$, some analysis gives that a bound on the error is
  \begin{align}\label{eqn::Error_Bound0}
    \|\boldsymbol{e}(t)\|^2 = |\boldsymbol{e}(t)^\dagger\boldsymbol{e}(t)| 
     \le \frac{4(d-1)}{\omega_a} \left[ 2\|g\|_{\infty} + t \left(\|g'\|_{\infty} + \|g\|_{\infty} ( C_1 + C_2)\right)\right],
  \end{align}
where $C_1 \equiv \left(\frac{\xi_a}{2}(d-1)(d-2) + 4 \|g\|_{\infty} \sqrt{d-1} \right)^2$ and $C_2 \equiv \left(\frac{\xi_a}{2}(d-1)(d-2)+ 2 \|g\|_{\infty} \sqrt{d-1} \right)^2$.
For a particular case let $g = \Omega$ be constant and suppose $d = 2$. Then
  \begin{align}\label{eqn::Error_Bound}
    \|\boldsymbol{e}(t)\|^2 \le \frac{8|\Omega|}{\omega_a}(1+10t|\Omega|^2),
  \end{align}
so that if $T$ is the final simulation time and $8|\Omega|(1+10T|\Omega|^2)/\omega_a < \epsilon^2$ then $\|e\| < \epsilon$.

Further, another (simpler) bound that can be derived is 
  \begin{align*}
     \|\boldsymbol{e}(t)\|^2 \le
     \left(4t(d-1)\|g\|^2_{\infty}\right)^2,
  \end{align*}
so that by Minkowski's inequality
  \begin{align*}
    \|e(t)\| \le 4t(d-1)\|g\|^2_{\infty} \le 4t(d-1)\left(\|p\|_{\infty} + \|q\|_{\infty} \right)^2.
  \end{align*}

For a simple example, we consider a single qubit system with $d = 2$ levels. The control functions $p = 1\text{e-}2$ and $q = 0$ are constant with corresponding (rotating frame) Hamiltonian 
  \begin{align*}
    \tilde{H}(t) = -i p(t) (a+a^\dag).
  \end{align*}
For each of the following experiments, the initial condition will be a qubit initialized to the ground state i.e. $\psib_0 = [1, 0]^T$. We consider three regimes: $\omega < p$, $\omega \approx p$, and $\omega > p$. We thus choose $\omega_1 = 2 \pi/10^4$, $\omega_2 = 2\pi/100$, $\omega_3 = 2 \pi$ with corresponding final times of $T_1 = 10^4$, $T_2 = 10^3$, $T_3 = 10^2$. We compute the forward problem for both the rotating frame transformation Hamiltonian and the rotating wave approximation Hamiltonian using the implicit midpoint rule and compute the error at each time gridpoint. Below we plot the squared norm of the error against the bound in \eqref{eqn::Error_Bound}.
\begin{figure}[H]
\graphicspath{{Plots/}}
\begin{center}
\includegraphics[width=0.47\textwidth]{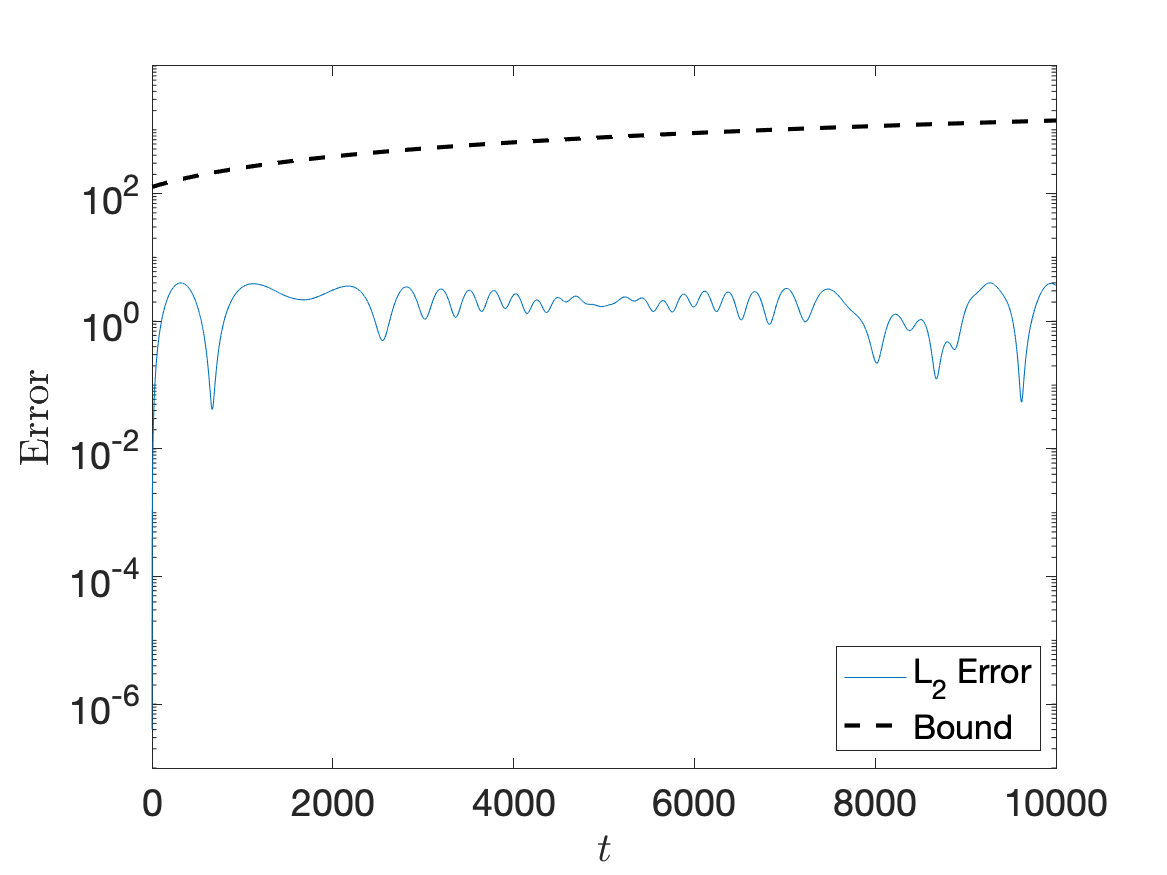}
\includegraphics[width=0.47\textwidth]{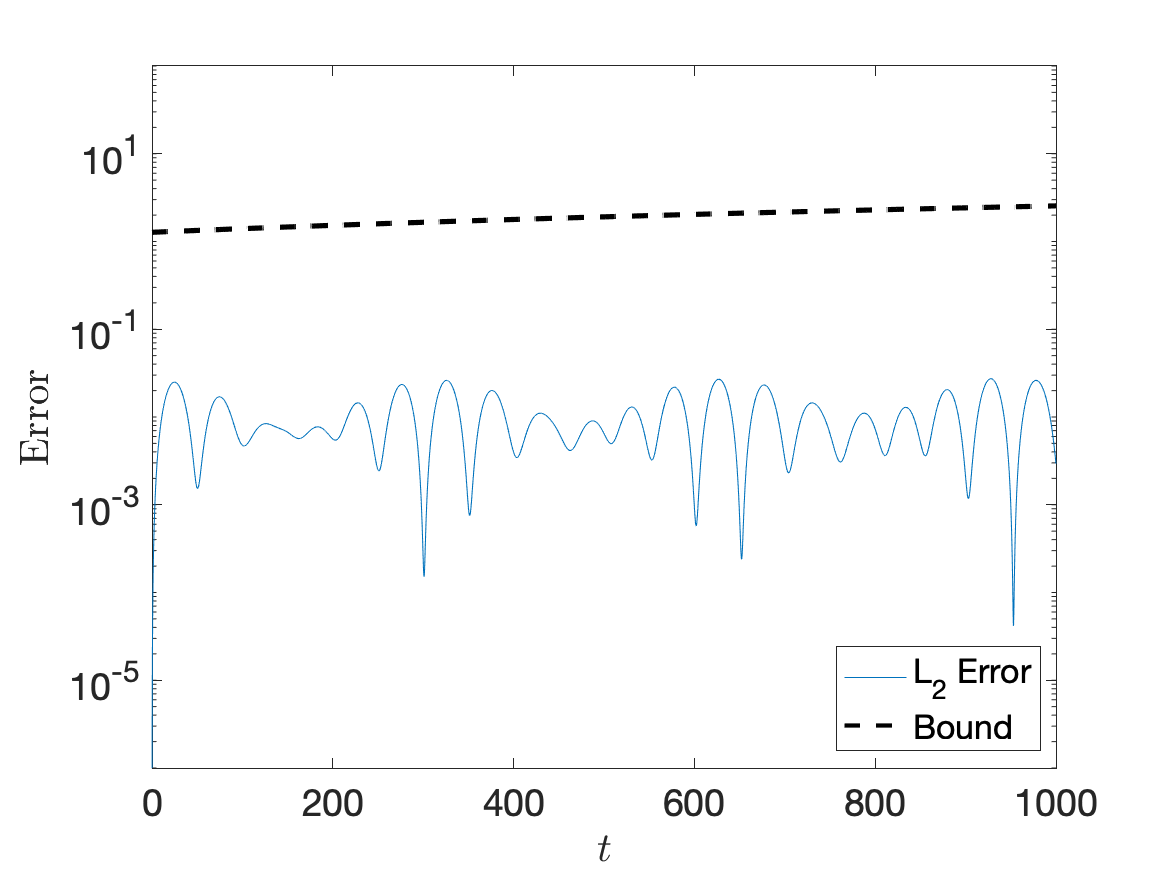}\\
\includegraphics[width=0.47\textwidth]{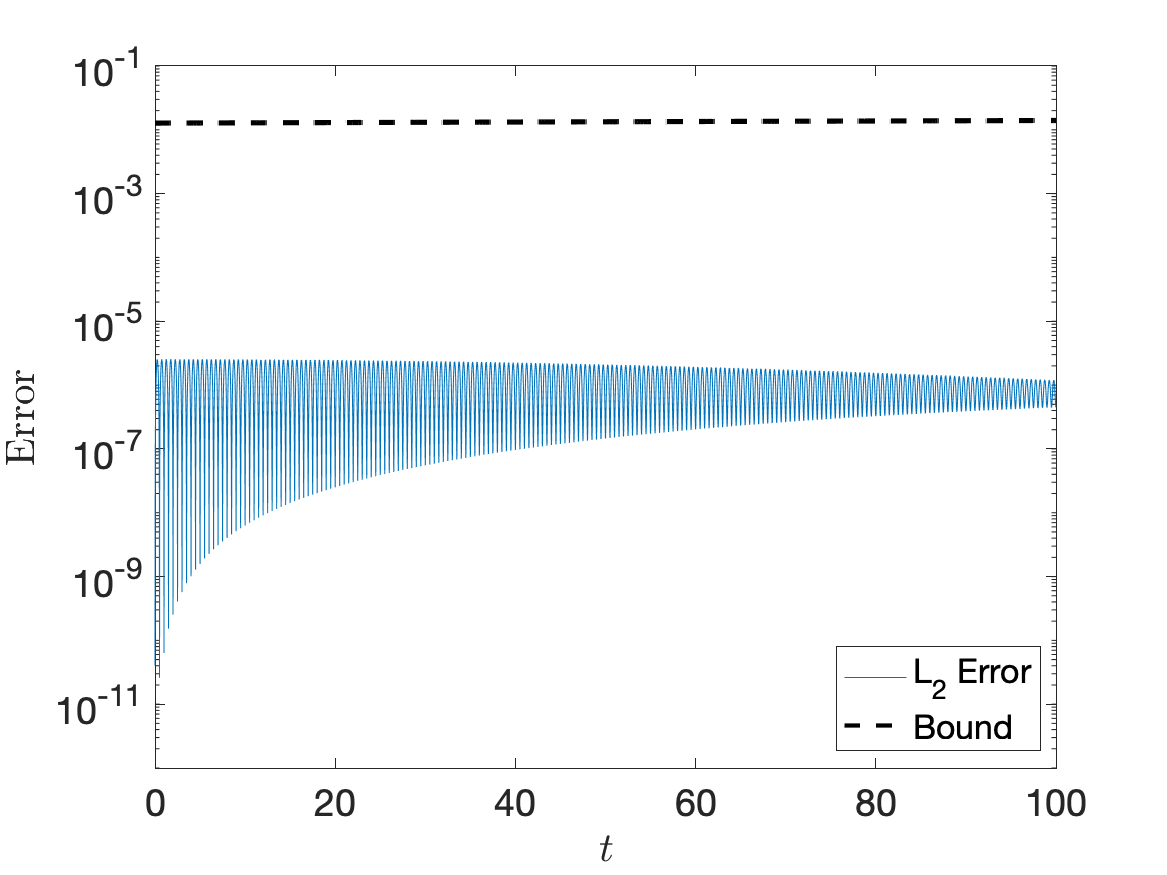}
\caption{Error of the rotating wave approximation for various choices of $\omega$. (Top left) $\omega = 2 \pi/10^4$, (Top right) $\omega = 2 \pi/10^2$, (Bottom) $\omega = 2 \pi$.  \label{fig:rwa_error}}
\end{center}
\end{figure}
We observe that in the regime where $\omega \le p$ the error bound is greater than one so that the bound is uninformative. In the extreme case where $\omega \ll p$ the error is larger than one and the approximation is quite poor, though it may still give some accuracy for frequencies close to the amplitude of the control signal. From the above it is clear that the approximation is more valid as $\omega$ grows larger than the control function amplitude, though the bound appears to be pessimistic.

\subsection{Unitary gate transformation}
A unitary gate transforms the quantum state $\psib_0$ into $\psib_1 = U \psib_0$, where $U$ is a unitary matrix. Also see \S~\ref{sec_unitary_evolution}.

\subsection{Trace fidelity of a unitary gate transformation}

The trace fidelity of a unitary gate transformation is a projective measure of the distance between two unitary matrices, $U$ and $V$, both belonging to $\mathbb{C}^{d\times d}$. The trace fidelity is defined by
\[
F(U,V) = \frac{1}{d^2}\left| (U,V)_F \right|^2 = \frac{1}{d^2}\left| \mbox{Tr}(U^\dagger V)\right|^2.
\]
It satisfies
\[
F(V,U) = F(U,V),\quad F(U,U) = 1,\quad 0\leq F(U,V) \leq 1.
\]
Note that the trace fidelity is invariant to the global phase difference between $U$ and $V$. For example,
\[
F(U, e^{i\theta}V) = \frac{1}{d^2} (U,e^{i\theta}V)_F  (e^{i\theta}V,U)_F 
\frac{1}{d^2} e^{i\theta} (U,V)_F  e^{-i\theta}(V,U)_F = F(U,V),
\]
for all phase angles $\theta \in \mathbb{R}$.

\newpage
\section{Open quantum systems}\label{sec_open}
In this section we are looking for definitions in terms of the density matrix, usually denoted by $\rho$.

\subsection{Open quantum system definition}
An open quantum system can interact with its environment and is by definition not closed. For example, let $A$ denote the quantum system of interest and let it be interacting with its surrounding, which often is called the bath. Together, the system and the bath can comprise the lab, or even the entire universe. The system $A$ could be a quantum computer, a molecule, or any other system we are interested in studying.

\subsection{Unitary evolution}
The combined quantum system (system of interest and the bath) is described by its density matrix. It evolves according to the Liouville-von Neumann equation, which is equivalent to the Schr\"odinger equation,
\[
\dot{\rho} = -i [H,\rho] = -i (H\rho - \rho H),
\]
where $H$ is the Hamiltonian.

\subsection{What does $|01\rangle\langle 10|$ mean?} \label{sec:outer_prod}
In Dirac notation, $|01\rangle$ is a basis vector in a composite system consisting of two sub-systems, $A$ and $B$. It is short hand for
\[
|01\rangle = |0\rangle_A \otimes |1\rangle_B = \eb^A_0 \otimes \eb^B_1.
\]
where $\eb^{A,B}_j$ denote the canonical basis vectors in systems $A$ and $B$, which can be of different dimensions. If both sub-systems have dimension two (qubits), the basis vector $01\rangle$ is a column vector of dimension four,
\[
|01\rangle = 
\begin{bmatrix} 1 \\ 0 \end{bmatrix}\otimes
\begin{bmatrix} 0 \\ 1 \end{bmatrix} =
\begin{bmatrix} 0 \\ 1 \\ 0 \\ 0\end{bmatrix}.
\]
The Hermitian conjugate of a basis vector is defined by
\[
\langle 10| = (|10\rangle)^\dagger = (\eb^A_1 \otimes \eb^B_0)^\dagger = (\eb^A_1)^\dagger \otimes (\eb^B_0)^\dagger.
\]
In the above case, $\langle 10|$ is the row vector of dimension four,
\[
\langle 10|=
\begin{bmatrix} 0 & 1 \end{bmatrix}\otimes
\begin{bmatrix} 1 & 0 \end{bmatrix}
=
\begin{bmatrix} 0 & 0 & 1 & 0 \end{bmatrix}.
\]
Thus, $|01\rangle\langle 10|$ is the outer product between $|01\rangle$ and $\langle 10|$. This is a square rank-1 matrix with one non-zero element. In the two by two case, the matrix has dimension $4\times 4$,
\[
|01\rangle\langle 10| = \begin{bmatrix} 0 \\ 1 \\ 0 \\ 0\end{bmatrix} 
\begin{bmatrix} 0 & 0 & 1 & 0 \end{bmatrix} =
\begin{bmatrix} 0 & 0 & 0 & 0\\
0 & 0 & 1 & 0 \\ 
0 & 0 & 0 & 0 \\ 
0 & 0 & 0 & 0
\end{bmatrix}
\]

\subsection{Pure state}
If a quantum system is in a pure state, its density matrix can be written as
\[
\rho = |\psi\rangle\langle \psi| = \psib \psib^\dagger,\quad \mbox{Tr}(\rho)=1,
\]
where $|\psi\rangle = \psib$ is the state vector of the system. Because the state vector is normalized, $\langle \psi|\psi\rangle = 1$, this implies
\[
\mbox{Tr}(\rho^2) = \mbox{Tr}(|\psi\rangle\langle \psi|\psi\rangle\langle \psi|) =
\mbox{Tr}(|\psi\rangle\langle \psi|) = \mbox{Tr}(\rho) = 1.
\]

\subsection{Mixed state}
A quantum system that is not in a pure state is said to be in a mixed state. In this case, its density matrix corresponds to a statistical ensemble of pure states, $\{p_k, |\psi_k\rangle\}_{k=1}^N$, where $p_k$ is the probability of the system being in the state $|\psi_k\rangle$. The density matrix corresponding to a mixed state can be written as
\[
\rho = \sum_{k=1}^N p_k |\psi_k\rangle \langle \psi_k|,\quad 0 \leq p_k\leq 1,\quad \sum_{k=1}^N p_k = 1.
\]

For a given density matrix, we can test if the quantum system is in a pure or a mixed state via a trace:
\[
\mbox{Tr}(\rho^2) = \begin{cases} 
1,& \mbox{pure},\\ 
< 1, & \mbox{mixed}.
\end{cases}
\]

\subsection{Purity}
Density matrices belong to the Hilbert-Schmidt space for linear operators $\rho\in\mathbb{C}^{d\times d}$ with Tr$(\rho)=1$, $\rho^\dagger = \rho$ and $\rho \geq 0$. This function space is endowed with the inner product $\langle A, B\rangle_{HS} = \mbox{Tr}(A^\dagger B)$ for any two operators $A$ and $B$ acting on the same Hilbert-Schmidt space. This inner product defines the length of a density matrix by
\[
\|\rho\|_{HS} = \sqrt{\langle \rho, \rho \rangle_{HS}} = \sqrt{P},
\]
where the quantity $P = \mbox{Tr}(\rho^2)$ is called the purity of a quantum system. The purity can thus be bounded by $1/d \leq P \leq 1$.

\subsection{Unitary freedom of ensemble} 
A density matrix for a quantum system might be
\[
\rho = \begin{bmatrix} 3/4 & 0\\ 0 & 1/4 \end{bmatrix} = \frac{3}{4} \begin{bmatrix} 1 & 0\\ 0 & 0 \end{bmatrix} + \frac{1}{4} \begin{bmatrix} 0 & 0\\ 0 & 1 \end{bmatrix}.
\]
As the eigenvectors  of the $2 \times 2$ identity matrix are  
\[
\eb_0 =  \begin{bmatrix} 1 \\ 0 \end{bmatrix}, \ \ \eb_1 = \begin{bmatrix} 0 \\ 1 \end{bmatrix},  
\]
we just expressed $\rho$ as 
\[
\rho = \frac{3}{4} \eb_0 \eb_0^T + \frac{1}{4} \eb_1 \eb_1^T.
\]

Equivalently (and perhaps surprisingly) we may express the same matrix $\rho$ as 

\[
\rho = \frac{1}{2}  \ab_+ \ab_+^T + \frac{1}{2} \ab_- \ab_-^T,
\]
where 
\[
\ab_{\pm} = \sqrt{\frac{3}{4}}  \eb_0 \pm \sqrt{\frac{1}{4}} \eb_1.
\]

Suppose we have two ensembles  corresponding to the columns of the matrices $E$ and $A$ then these only generate the same density matrix iff there is a unitary matrix $U$ such that $E=UA$.

\subsection{Commutator}
The commutator symbol $[\cdot,\cdot]$ is shorthand notation for the specific matrix operation
\[
[H,\rho] \equiv H\rho - \rho H.
\]
\YC{Note that if $H$ and $\rho$ happen to commute, then $[H,\rho] = 0$.}
\subsection{Anti-commutator}
As with the commutator, the anti-commutator is a shorthand notation: 
\[
\{H,\rho\} \equiv H\rho + \rho H.
\]
\subsection{Hilbert-Schmidt scalar product}
The Hilbert-Schmidt inner (scalar) product is $\langle A, B\rangle_{HS} = \mbox{Tr}(A^\dagger B)$ for any two operators $A$ and $B$ acting on the same Hilbert-Schmidt space.
\subsection{Measurement outcome}
Quantum measurements are described by a set $\{M_k\}_{k=1}^N$ of measurement operators satisfying the constraint $\sum_k M_k^{\dagger} M_k = I$. Given a state $ \psib \in \mathcal{H}$, instantaneously after the measurement it becomes
\begin{align}
    \psib \rightarrow
    \frac{M_k \psib }{ \sqrt{p_k}},
    \quad
    p_k = \psib^\dag M_k^{\dagger} M_k \psib = \| M_k  \psib \|^2 \geq 0.
\end{align}
The measurement outcome is the index $k$ of the state that resulted.
\subsection{Measurement transformation}
Referring back to \eqref{eq_m-trans} we have an expression for the transformation of a measurement operator $M_k$ on the state $\psib$. Using the definition of the density matrix we have
    \begin{align*}
        \rho \rightarrow \psib_k \psib_k^\dagger = \left(\frac{M_k \psib}{\sqrt{p_k}}\right)\left(\frac{ \psib^\dag M_k^\dag}{\sqrt{p_k}}\right) = \frac{M_k \rho M_k^\dag}{p_k},
    \end{align*}
with probability $p_k$.

\subsection{Observable}

An observable $A\in \mathbb{C}^{d\times d}$ is a Hermitian operator, $A^\dagger = A$. From the spectral theorem, an observable can be decomposed as
\[
A = \sum_{k=0}^{d-1} \lambda_k \ab_k \ab_k^\dagger,
\]
where $\lambda_k$ and $\ab_k$ are the eigenvalues and corresponding orthonormal eigenvectors of $A$. The measurement operators for an observable are therefore projection operators,
\[
P_k = \ab_k \ab_k^\dagger.
\]
The projectors are orthonormal, $P_k P_l = \delta_{kl} P_l$.

\subsection{Expectation of an observable}
Consider an observable $O$ measured for a system in the pure state ensemble $\{p_i, \psib_i\}$. Using the definition of the expectation of an observable in terms of the state vector of a pure quantum system \eqref{eqn::Expectation}, the expectation for the pure state ensamble becomes 
    \begin{align*}
        \langle O\rangle_\rho = \sum_i p_i \psib_i^\dag O \psib_i = \sum_i p_i \text{Tr}(O\psib_i \psib_i^\dag) = \text{Tr}(O\sum_i p_i \psib_i \psib_i^\dag) = \text{Tr}(O\rho).
    \end{align*}

\subsection{Partial trace}
Consider an operator $O$ acting on a bipartite system $H = H_A \otimes H_B$. Let $\{\psib\}$ denote basis vectors that span the sub-system space $H_B$. The partial trace with respect to $H_B$ is defined as
\begin{align}
    \mbox{Tr}_B(O) &:= \sum_{\psib} \langle\psib| O |\psib\rangle \\
     &= \sum_{\psib} (I_A \otimes \psib^\dagger) O (I_A \otimes \psib)
\end{align}
where it is understood in the first line that $\psib$ acts only on the second Hilbert space $H_B$. The resulting operator acts on the Hilbert space $H_A$.

Note, that if $O = M_A \otimes N_B$, then 
\begin{align}
    \mbox{Tr}_B(M_A \otimes N_B) = M_A \mbox{Tr}(N_B)
\end{align}

\subsection{Reduced density matrix}
Consider a bipartite system with Hilbert space $H = H_A \otimes H_B$, where the dimensions of $H_A, H_B$ are $d_A, d_B$ so that $d_A d_B = \dim(H)$. Let $\rho$ be the density matrix of the combined system. The reduced density matrix $\rho_A$ is the density matrix corresponding to the subsystem $H_A$ only. It is given by taking the partial trace with respect to $H_B$:
\begin{align}
    \rho_A := \mbox{Tr}_B(\rho).
\end{align}

Generally, for given orthonormal basis vectors $\{\psib_i\}_{i=0}^{d_A-1}$ in $H_A$ and $\{\phib_k\}_{k=0}^{d_B-1}$ in $H_B$, the density matrix $\rho$ acting on $H$ can be written in this basis as
\begin{align}
    \rho = \sum_{ijkl} \lambda_{ijkl} \psib_i \psib_j^\dagger \otimes \phib_k \phib_l^\dagger,\quad \mbox{Tr}(\rho)=1,
\end{align}
for coefficients $\lambda_{ijkl} \in \C$. The reduced density matrix $\rho_A$ is thus a contraction over the two last indices
\begin{align}
    \rho_A = \mbox{Tr}_B(\rho) = \sum_{i,j} \left( \sum_m \lambda_{ijmm} \right) \psib_i \psib_j^\dagger,\quad \mbox{Tr}(\rho_A)=1.
\end{align}

\subsection{Entangled and separable states}
A composite quantum system consisting of two or more subsystem can be in an entangled or separable state. In the case of two subsystems (A and B), the density matrix for a separable state can be written as 
\[
\rho = \sum_i p_i \left(\rho_i^A \otimes \rho_i^B\right),
\]
where $0\leq p_i\leq 1$ are the probabilities of systems A and B to be in the states $\rho_i^A$ and $\rho_i^B$, respectively.
A system that is not separable is entangled.

Note: The distinction between separable and entangled states is different from the distinction between pure and mixed states. 

\subsection{Maximally mixed state}
A density matrix $\rho\in\mathbb{C}^{d\times d}$ is said to be in a maximally mixed state if its purity, $P=\mbox{Tr}(\rho^2) = 1/d$. The probabilities in the corresponding ensamble of pure states, $\{p_j, \Psi_j\}_{j=1}^d$, satisfy $p_j=1/d$ for all $j$. 
If the state vectors in the pure state ensamble are orthonormal, they form a basis in which the corresponding density matrix satisfies $\rho = d^{-1} I$.

\subsection{Maximally entangled state}\label{sec_max-entangled}
Two coupled quantum subsystems (A and B) are maximally entangled if the reduced density matrix $\rho_A = \mbox{Tr}_B(\rho)$ is maximally mixed. 

Does this imply that also $\rho_B = \mbox{Tr}_A(\rho)$ is maximally mixed?

\subsection{Observable on subspace}
Given an observable $M_A$ acting on a subsystem $A$ of a bipartite system with Hilbert space $H=H_A \otimes H_B$, the expected value of that observable is given by
\begin{align}
    \langle M_A \rangle = \mbox{Tr}(M_A \rho_A) = \mbox{Tr}\left( \left(M_A \otimes I_B \right)\rho \right) = \langle M_A \otimes I_B \rangle,
\end{align}
where $\rho_A$ denotes the reduced density matrix, and $I_B$ is the identity matrix in subsystem $B$.

\subsection{Liouville-von Neumann equation}
Consider a single component system and recall that the state vector $\psib$ evolves according to the Schr\"odinger equation. By definition $\rho(t) = \psib \psib^\dag$, so that using product rule
    \begin{align}\label{eqn::LvN}
        \frac{d}{dt} \rho(t) 
        = \dot{\psib} \psib^\dag + \psib \dot{\psib}^\dag 
        = -iH(t)\psib \psib^\dag + \psib (i \psib^\dag H(t)) 
        & = -iH(t) \rho(t) - (-i \rho(t) H(t)) \nonumber \\
        & = -i[H,\rho],
    \end{align}
which we see is a generalization of Schr\"odinger's equation to density matrices and is called the \textit{Liouville-von Neumann} equation. Note that if we write this in the superoperator form (i.e. vectorizing $\rho$) we get 
    \begin{align*}
        \frac{d}{dt} \text{vec}(\rho) = - i(H(t) \otimes I - I \otimes H(t)) \text{vec}(\rho) = -i \hat{H}(t) \text{vec}(\rho).
    \end{align*}
$-i\hat{H}$ is skew-Hermitian which, like in the Schr\"odinger case, implies that the evolution of this equation is unitary.

\subsection{Lindblad master equation}
In a closed system the evolution of the state must be unitary. In open systems, however, this may no longer be the case. We can modify the Liouville-von Neumann equation \eqref{eqn::LvN} by adding damping terms to capture the dynamics of an open system. We then consider the equation
    \begin{align}\label{eqn::Lindblad}
        \frac{d}{dt} \rho(t) 
         = -i[H,\rho] + \mathcal{L}_D \rho = \mathcal{L} \rho(t),
    \end{align}
where $\mathcal{L}_D$ has dissipative terms and if it is non-zero we have decoherence. The generator of the evolution, $\mathcal{L}$, is called the \textit{Lindbladian}. The specific form of the dissipative term is
    \begin{align}\label{eqn::LindbladDecay}
        \mathcal{L}_D \rho(t) = \sum_{\alpha} \gamma_\alpha \left(L_\alpha \rho(t) L_\alpha^
        \dag - \frac{1}{2}\left(L_\alpha^{\YC{\dag}} L_\alpha^{\YC{ }} \rho(t) + \rho(t)L_\alpha^{\YC{\dag}} L_\alpha^{\YC{}}   \right)\right),
    \end{align}
where we often consider Lindblad operators that model decay, $L_\alpha = a$, and dephasing, $L_\alpha = a^\dag a$. The coefficients $\gamma_\alpha$ represent inverse half-lifes for the corresponding decay process so that necessarily $\gamma_\alpha > 0$.

\subsection{Collapse operator}
To define the Linblad master equation \eqref{eqn::Lindblad} we required the addition of a dissipative term $\mathcal{L}_D$. The operators $\sqrt{\gamma_\alpha} L_\alpha$ found in the dissipative term \eqref{eqn::LindbladDecay} are called \textit{collapse operators} as they encapsulate dissipative processes.

\subsection{Factorized state}
Consider a two component system, $H = H_A \otimes H_B$. We say $\rho$ is a factorized state if we may write it as a tensor product of density matrices for each system, i.e. $\rho(t) = \rho_A(t) \otimes \rho_B(t)$. This result can be generalized to an arbitrary number of component systems. We note that writing the density matrix of the full system in this way is not always possible and only occurs when the systems have no correlation and are decoupled.

\subsection{Kraus operator sum representation (OSR)}
Consider a quantum system $\rho_S$ combined with a bath $\rho_B$ with separable initial state $\rho_{SB}(0) = \rho_S(0)\otimes \rho_B(0)$. Let $\{\mu\}$ denote an orthonormal basis of the bath eigenstates, and let $\{\nub\}, \{\lambda_{\nub}\}$ form an orthonormal spectral decomposition of the bath at $t=0$, i.e. $\rho_B(0) = \sum_{\nub} \lambda_{\nub} \nub \nub^\dag$. 

The evolution of the open quantum system $\rho_S(t)$ can then be written in terms of the Kraus operator sum representation (OSR)
\begin{align}
    \rho_S(t) = \sum_{\mub,\nub} K_{\mub \nub}(t) \rho_S(0) K^\dag_{\mub \nub}(t),
\end{align}
where the Kraus operators $K_{\mub \nub}(t)$ are given by 
\begin{align}\label{eq:OSR}
    K_{\mub \nub}(t) = \sqrt{\lambda_{\nub}} \left(I_S \otimes \mub^\dag\right) U(t) \left( I_S \otimes \nub\right)
\end{align}
and $U(t) = e^{-iHt}$. 

The OSR can be derived considering the joint evolution of the system and the bath, $\rho_{SB}(t)$, which is unitary (solving Schr\"odinger's equations, and is hence given by $\rho_{SB}(t) = U(t) \rho_{SB}(0) U^\dag(t)$ where $U(t) = e^{-iHt}$. The OSR follows from taking the partial trace over the bath in the orthonormal basis of bath eigenstates $\{\mub\}$.

Note that the OSR defines a linear mapping from an initial state $\rho(0)$ to any state in time $\rho(t)$. 

\subsection{Evolution under a single Kraus operator}
The Kraus operator sum representation (OSR) is more general than the Schr\"odinger equation, as it contains the latter as a special case where $U(t) = U_S(t)\otimes U_B(t)$. In this case, we have $K_{\mub\nub} = U_S \sqrt{\lambda_{\nub}} \mub^\dag U_B \nub$, and it follows that
\begin{align}
    \rho_S(t) = \sum_{\mub,\nub} K_{\mub\nub}(t) \rho_S(0) K^\dag_{\mub\nub}(t) = U_S(t) \rho_S(0) U^\dagger_S(t).
\end{align}
Here, $U_S(t)$ is considered a \textit{single} Kraus operator.

\subsection{Non-selective measurements}
Let us observe that an OSR represents more than dynamics, it can also capture measurements. Specifically, consider measurement operators $\{ M_k\}$ with $\sum_k M^{\dagger}_k M_k =I$. Recall that a state subjected to this measurement maps to
\begin{equation}
\rho\mapstochar \rightarrow \rho_k = \frac{M_k \rho M^{\dagger}_k}{\text{Tr}[ M_k \rho M^{\dagger}_k]},
\end{equation}
with probability $\text{Tr}[ M_k \rho M^{\dagger}_k]$. Consider the case where we perform this measurement but do not learn the outcome $k$. What happens to $\rho$ after this measurement? In this case  
\begin{equation}
    \rho\mapstochar \rightarrow \langle \rho \rangle = \sum_k p_k \rho_k = \sum_k M_k \rho M^{\dagger}_k, 
\end{equation}
which we recognize as a non-selective measurement and is in Kraus OSR form.  

\subsection{Positive operators, maps and flow}
An operator $A$ is positive if all its eigenvalues are non-negative, but not all of them are zero. Note that all positive operators are positive semi-definite, but not all positive semi-definite operators are positive. A density operator is a positive operator because it is defined from an ensemble of pure states with non-negative probabilities that sum to one.

A map (also known as a super-operator, process, or channel) transforms an operator into another operator. The state of a quantum system can be described by a time-dependent density operator, $\rho(t)$. Thus, the transformation from the initial density operator to the final density operator can be expressed by a map $\Phi$:
\[
\rho(t) = \Phi[\rho(0)],\quad \Leftrightarrow\quad \Phi:\rho(0)
\mapsto \rho(t)
\]
A map $\Phi$ is said to be positive if it transforms positive operators into positive operators.

In the context of systems of ordinary differential equations (ODEs), the solution operator map is often called the flow. For a Hamiltonian system of ODEs, the flow is area preserving. A numerical time integration method that is area preserving is called symplectic.

\subsection{Completely positive map}
Let $\mathcal{K}_R$ be an ancillary Hilbert space of dimension $k$. The map $\Phi$ is a completely positive (CP) map if $\Phi$ is a positive map and $\Phi\otimes \mathcal{I}_R^{(k)}$ is also a positive map, for all natural numbers $k$. Here, $\mathcal{I}_R^{(k)}$ is the identity map on $\mathcal{K}_R$.

\subsection{Partial transpose}
Let $\mathcal{I}_r$ be the $r\times r$ identity map, where $r=1,2,3,\ldots$. To check if a map $T$ is completely positive, we need to check if an extension $T^p = T \otimes \mathcal{I}_r$ of $T$ is also positive. This extension is called the partial transpose, and its action on any basis element of $\mathcal{B}( \mathcal{H}_S \otimes \mathcal{H}_R)$ is as follows:
\begin{equation}
    T^p (|i \rangle\langle j| \otimes |\mu \rangle \langle \nu |) = |j \rangle \langle i| \otimes | \mu \rangle \langle \nu |.
\end{equation}

\subsection{Quantum map}
We define a quantum map (or quantum channel) as a map that is (1) trace preserving, (2) linear, and (3) completely
positive. This definition is motivated by the fact that we know that such maps have a Kraus OSR, and that the Kraus OSR arises both from the physical prescription of unitary evolution followed by partial trace, and from (non-selective) measurements.

\subsection{Negative partial transpose and entanglement}

Consider a separable (thus by definition un-entangled) state $\rho = \sum_i p_i \rho_i^A \otimes \rho_i^B, $ where $p_i$ are the probabilities with $\rho_i^A$ and $\rho_i^B$ being the corresponding quantum states (positive, normalized). The state $\rho$ obviously arises from the mixed state ensemble $\{p_i, \rho_i^A \otimes \rho_i^B\}$, in which every element is a tensor product state. Mixing such states classically does not generate any entanglement between A and B, hence the definition. 

Applying the partial transpose yields:

\begin{equation}
    T^p(\rho) = (T \otimes \mathcal{I})(\rho) = \sum_i p_i T(\rho_i^A) \otimes \rho_i^B = \sum_i p_i \sigma_i^A \otimes \rho_i^B.
\end{equation}
Since the transpose does not change the eigenvalues, $\sigma_i^A = T(\rho_i^A)$ is a valid quantum state, and hence $T^p(\rho)$ is another separable state. In particular, this shows that every separable state has a positive partial transpose (PPT). In other words separability implies PPT. Conversely, a negative partial transpose (NPT) implies entanglement. This means that PPT is a necessary condition for separability. Conversely, a negative partial transpose (NPT) implies entanglement. This means that PPT is a necesssary condition for separability. 

\subsection{Concurrence}
Consider a pure state $\psib$ in the tensor product space $\mathcal{H}_A \otimes \mathcal{H}_B$ of finite dimensional Hilbert spaces $\mathcal{H}_A, \mathcal{H}_B$ for two systems $A$ and $B$. The concurrence is defined by
  \begin{align*}
    C(\psib) = \sqrt{2(1 - \text{Tr} \rho^2_A)},
  \end{align*}
where $\rho_A$ is the (usual) reduced density matrix obtained by tracing over system $B$. We note that in the case of a pure state the concurrence gives a value of zero, and that $0 \le C \le \sqrt{2}$. We may extend the above definition to mixed states $\rho$ by considering the convex roof
  \begin{align*}
    C(\rho) \equiv \min_{\{p_k, \psib_k\}} \sum_k p_k C(\psib_k),
  \end{align*}
for all possible ensemble realizations $\rho = \sum_k p_k \psib_k \psib_k^\dag$, where $p_k \ge 0$ and $\sum_k p_k = 1$. It follows that a state $\rho$ is \textit{separable} if and only if $C(\rho) = 0$ so that the concurrence gives a measure of the amount of entanglement in a given quantum state.

\subsection{Fidelity between arbitrary states}

The fidelity between two arbitrary quantum states, described by their respective density matrices $\rho$ and $\sigma$, is defined by
\begin{align}\label{eq_fid-arb}
    F(\rho, \sigma) = \mbox{tr}\left(\sqrt{\rho^{1/2} \sigma \rho^{1/2}}\right).
\end{align}

\subsection{Fidelity between a pure and an arbitrary state}

Let the state vector of the pure state be $|\psi\rangle$, corresponding to the density matrix $\rho = |\psi\rangle \langle\psi|$. We first note that $\rho$ is idempotent,
\begin{align*}
    \rho^2 = |\psi\rangle \langle\psi |\psi\rangle \langle\psi| = |\psi\rangle \langle\psi| = \rho.
\end{align*}
Thus, $\rho^{1/2} = \rho$. By construction, $\rho$ is a rank one matrix with one non-zero eigenvalue,
\begin{align*}
    \rho |\psi\rangle = \lambda |\psi\rangle,\quad \lambda=1.
\end{align*}
All other eigenvalues of $\rho$ are zero. Since the trace of a matrix equals the sum of its eigenvalues,
\begin{align*}
    \mbox{tr}(\rho) = \mbox{tr}\left(|\psi\rangle \langle\psi| \right) = 1.
\end{align*}
Because $\rho^{1/2} = \rho = |\psi\rangle \langle\psi|$, the definition \eqref{eq_fid-arb} of the fidelity between arbitrary states gives
\begin{align}
    F(|\psi\rangle \langle\psi|, \sigma) &= 
    \mbox{tr}\left(\sqrt{|\psi\rangle \langle\psi| \sigma |\psi\rangle \langle\psi|}\right)\nonumber \\
    &= \sqrt{\langle\psi| \sigma |\psi\rangle}\, \mbox{tr}\left(\sqrt{|\psi\rangle \langle\psi|}\right)\nonumber \\
    &= \sqrt{\langle\psi| \sigma |\psi\rangle} \, \mbox{tr}\left(|\psi\rangle \langle\psi|\right) = 
    \sqrt{\langle\psi| \sigma |\psi\rangle},\label{eq_fid-pure-arb}
\end{align}
where the second equality follows because $\langle\psi| \sigma |\psi\rangle$ is a scalar. The third and fourth equality's follow because $|\psi\rangle\langle \psi|$ is idempotent with trace one.

If both states are pure, we can write $\sigma = |\phi\rangle \langle \phi|$ for some state vector $|\phi\rangle$. The formula \eqref{eq_fid-pure-arb} becomes
\begin{align}
    F(|\psi\rangle \langle\psi|, |\phi\rangle \langle\phi|) = 
    \sqrt{\langle\psi| \phi\rangle \langle\phi |\psi\rangle} = | \langle\psi| \phi\rangle |.
\end{align}

\section{Acknowledgements}
We gratefully acknowledge financial support from the Laboratory Directed Research and Development (LDRD) program at Lawrence Livermore National Laboratory, grant 20-ERD-028, as well as financial support from the Advanced Scientific Computing Research (ASCR) program at DOE under the ARQC/TEAM project, grant 2019-LLNL-SCW-1683.

This work was performed under the auspices of the U.S. Department of Energy by Lawrence Livermore National Laboratory under Contract DE-AC52-07NA27344. This is contribution LLNL-TR-817270.

\bibliographystyle{plain}
\bibliography{refs}

\end{document}